\documentclass[aps,twocolumn,pra,floatfix,showpacs,lengthcheck,citeautoscript,superscriptaddress]{revtex4-1}
\usepackage{amsmath}
\usepackage{amssymb}
\usepackage{graphicx}
\usepackage{bm}
\usepackage{color}
\usepackage{times}
\usepackage{verbatim}
\usepackage[none]{hyphenat}

\newcommand{\bea}{\begin{eqnarray}}
\newcommand{\eea}{\end{eqnarray}}
\newcommand{\be}{\begin{equation}}
\newcommand{\ee}{\end{equation}}

\begin{document}
\title{Floquet Spectrum and Two-Terminal Conductance of a Transition Metal Dichalcogenide Ribbon under a Circularly Polarized Laser Field.}
\author{A. Huam{\'{a}}n}
\affiliation{Centro At{\'{o}}mico Bariloche,
Comisi\'on Nacional de Energ\'{\i}a At\'omica, 8400 Bariloche, Argentina}
\affiliation{Instituto Balseiro, Universidad Nacional de Cuyo, 8400 Bariloche, Argentina}
\affiliation{Consejo Nacional de Investigaciones Cient\'{\i}ficas y T\'ecnicas (CONICET), Argentina}
\author{Gonzalo Usaj}
\affiliation{Centro At{\'{o}}mico Bariloche,
Comisi\'on Nacional de Energ\'{\i}a At\'omica, 8400 Bariloche, Argentina}
\affiliation{Instituto Balseiro, Universidad Nacional de Cuyo, 8400 Bariloche, Argentina}
\affiliation{Consejo Nacional de Investigaciones Cient\'{\i}ficas y T\'ecnicas (CONICET), Argentina}

%%%%%%%%%%%%%%%%%%%%%%%%%%%%%%%%%%%%%%%%%%
\begin{abstract}
We study the transport properties of a monolayer transition metal dichalcogenide (TMDC) ribbon subject to a time periodic circularly polarized laser field. First, we calculate the quasienergy  spectrum within the framework of the Floquet theory and analyze the nontrivial topology of the Floquet bulk gaps. The later is revealed by the presence of chiral edge states inside the bulk gaps in finite samples, in agreement with the calculation of the appropriate winding numbers as a function of both the energy  and the amplitude of the laser field. The effect of the time dependent perturbation on the equilibrium edge states is also analyzed. Finally, we calculate the two-terminal conductance and discuss how the above mentioned effects manifest on it. In particular, besides the expected suppression of the bulk conductance and the emergence of edge transport at the Floquet gaps, we find that there is an asymmetry between left and right transmission coefficients (in the zigzag case), leading to pumping effects. In addition, we found that the laser field can lead to a complete switch off of the linear conductance when the later is dominated by the equilibrium edge states. 
\end{abstract}
%%%%%%%%%%%%%%%%%%%%%%%%%%%%%%%%%%%%%
\date{\today}
\maketitle

%%%%%%%%%%%%%%%%%%%%%%%%%%%%%%%%%%%%%%%%%%%%%%%%%%%%%%%%%%
\section{Introduction}\label{introduction}%%%%%%%%%%%%%%%%
%%%%%%%%%%%%%%%%%%%%%%%%%%%%%%%%%%%%%%%%%%%%%%%%%%%%%%%%%%
Much of attention in Condensed Matter Physics have been devoted in recent years to the study of new classes of atomically thin materials with a variety of very promising electronic and mechanical properties~\cite{Manzeli2017}. Monolayers of transition metal dichalcogenides (TMDC) are one family of such materials, characterized by the chemical formulae MX$_2$, where M is a transition metal atom---usually tungsten (W) or molybdenum (Mo)---, and X is a chalcogen atom---typically sulphur (S), selenium (Se) or tellurium  (Te)---, and a two dimensional crystal structure that corresponds to an hexagonal lattice as in the case of graphene~\cite{CastroNeto2009b}. 
One important difference with the later, however, is that these materials present a direct gap of a few eV, what makes them particularly interesting for fabricating semiconductor devices~\cite{li2014a,roy2014} and for photonics and optoelectronics applications~\cite{mak2016,wang2012c}. 

At the same time, there has also been much interest in the so-called topological materials~\cite{Kane2005,Koenig2007,Hsieh2008,Hasan2010,Ando2013} (a bulk insulator with conducting surface states) and in different ways of inducing and controlling their topological properties.
Among several proposals there was the idea that out of equilibrium systems might present new topological properties that are absent at equilibrium~\cite{Oka2009,Kitagawa2010,Lindner2011,Rudner2013}---for instance by applying a circularly polarized laser field on graphene~\cite{Gu2011,Zhou2011,Kitagawa2011,Calvo2011,Iurov2012,Perez-Piskunow2014,Usaj2014a,FoaTorres2014}---. This has opened a new research area of the now called Floquet topological insulators due to the method used in their investigation (Floquet theory \cite{Shirley1965,Sambe1973,Grifoni1998a,Kohler2005}). 
Since then, such nonequilibrium properties have been intensively investigated  in a variety of systems with the focus in many different aspects of the problem~\cite{Dora2012,Wang2013a,Rechtsman2013,Ezawa2013,Goldman2014,Choudhury2014,DAlessio2014,Dehghani2014,Liu2014,Fregoso2014,Kundu2014,Goldman2015,
Calvo2015,Nathan2015,DalLago2015,Carpentier2015,Perez-Piskunow2015,Seetharam2015,Iadecola2015,Dehghani2015,Sentef2015,Titum2015a,Farrell2015,Lovey2016,PeraltaGavensky2016,Lindner2017,Kundu2017,PeraltaGavensky2018a}.

An experimental confirmation of the existence of protected edge states in Floquet topological insulators has been recently achieved~\cite{Rechtsman2013}. In addition, the Floquet induced gaps have already been observed at the surface of a topological insulator by using time and angle resolved photoemission spectroscopy~\cite{Wang2013a} and more recently, effective Floquet Hamiltonians were realized in cold matter systems~\cite{Jotzu2014}. Hence, there is a clear need for the investigation of other possible scenarios for the observation of signatures of such phenomena~\cite{PeraltaGavensky2018}.

The Floquet approach has been applied to TMDCs mainly by using first principle techniques and tight-binding models to investigate the properties of the Floquet spectrum~\cite{Claassen2016,Giovannini2016}. In\-te\-res\-tin\-gly, the lack of inversion symmetry in monolayer TMDC dictates that optical transition between valence and conduction bands at the $K$ and $K'$ points of the Brillouin zone (BZ) are selectively forbidden when using left or right handed circularly polarized light~\cite{Xiao2012,Yao2008,Ezawa2012}. This asymmetry has been exploited in the realization of valley polarization~\cite{Cao2012,Zeng2012}, the experimental measurement of the optical Stark shift~\cite{Sie2014} and the generation of valley- and spin-polarized currents~\cite{Zhang2014,Hsieh2018}---the latter thanks to the relatively strong spin-orbit coupling in the valence band.

The band structure of generic TMDC has been studied using Density Functional Theory (DFT) calculations \cite{liu2013,cappelluti2013} and the results were used to fit different tight binding models (differing in the number of orbitals involved)~\cite{cappelluti2013}. A first approximation to this was a two-band model with orbitals $d_{z^2}$ and $d_{xy}\pm i d_{x^2}$  of the transition metal (the sign depending on the $K$ point), a model inspired in a graphene tight binding Hamiltonian with a mass term, which opens a gap as a result of the broken inversion symmetry~\cite{Xiao2012}. This model suffices for some calculations of valley and spin conductances~\cite{Xiao2012,Xiao2007,Yao2008}, although it has been shown to be insufficient when describing phenomena such as the optical Stark shift~\cite{Claassen2016}. This has led to proposals where more than two transition metal orbitals and even a contribution from $p$-orbitals from the chalcogen atoms are considered~\cite{cappelluti2013,kormanyos2015}.

In this work we use the three-band model for TMDC developed by Liu {\it et al}~\cite{liu2013} to describe the bands of a monolayer TMDC ribbons. The effect of the laser radiation normally incident upon the monolayer is included by using the usual Peierls substitution, resulting in a  time dependent Hamiltonian. The solution of the problem is obtained by using Floquet theory \cite{Shirley1965,Sambe1973,Grifoni1998a,Kohler2005,Rudner2013}. With the bulk Floquet Hamiltonian we calculate the topological invariants that explain the presence of chiral edge states in the Floquet spectrum. We also analyze the effect that the time dependent perturbation has on the equilibrium edge states that appear inside the bulk gap in finite width ribbons. Finally, we perform calculations of the zero temperature conductance in a two-terminal configuration with the aim to determine how different aspect of the Floquet spectrum manifest on such a physical observable. We find that, as expected, there is a suppression of the bulk conductance for electrons with an incident energy on the range of the Floquet gaps, coincident with the presence of edge transport. Additionally, we find that there is an asymmetry between left and right transmission coefficients for the case of zigzag ribbons, that leads to the appearance of pumping effects.  

This work is organized as follows. In Sec.~\ref{threeband} we describe the equilibrium band structure and the main ingredients of the TB model~\cite{liu2013}. In Sec.~\ref{floquetstates} we introduce the time dependent perturbation together with a brief discussion of some relevant aspects of the Floquet theory. The TMDC's Floquet spectrum, for both bulk and ribbons, is analyzed here and the appearance of chiral edge states is discussed by using the appropriate  topological invariants. Finally, we present our results for the two-terminal conductance in irradiated ribbons in Sec.~\ref{twoterminal}. A summary is given in Sec.~\ref{conclusions}. 
%%%%%%%%%%%%%%%%%%%%%%%%%%%%%%%%%%%%%
\begin{figure}[t]
\begin{center}
 \includegraphics[width=0.85\columnwidth]{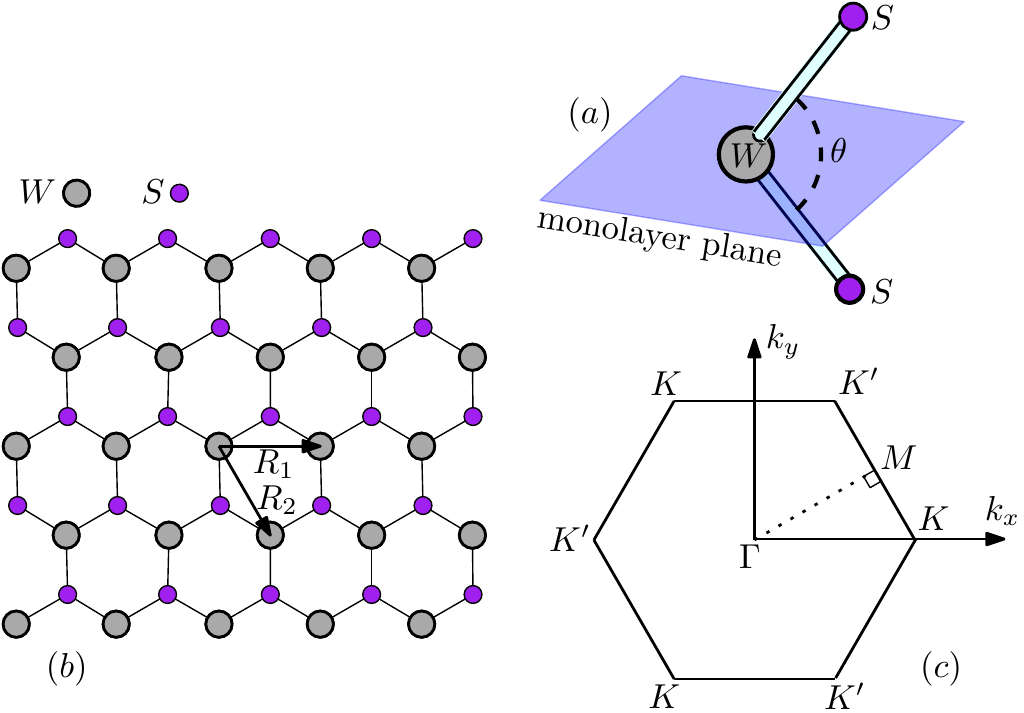}
 \caption{(Color online) (a) Crystal basis of a generic monoloyer TMDC (WS$_2$ in particular), showing a W atom (in plane) and two S atoms (out of plane). The two W-S links  form an angle $\theta\sim100^\circ$. (b)  Crystal Structure of a monolayer TMDC. The gray and purple circles represent the W and S atoms, respectively.  The tungsten atoms are arranged in an hexagonal Bravais lattice. (c) Brillouin zone exhibiting the two inequivalent high-symmetry points $K$ and $K'$ \label{1}.}
 \end{center}
\end{figure}
%%%%%%%%%%%%%%%%%%%%%%%%%%%%%%%%%%%%%%

%%%%%%%%%%%%%%%%%%%%%%%%%%%%%%%%%%%%%%%%%%%%%%%%%%%%%%%%%%%%%%%%%%%%%%%%%
\section{Three-Band Tight Binding Model with Third Nearest Neighbors\label{threeband}}
\subsection{Model and Symmetry Considerations}
%%%%%%%%%%%%%%%%%%%%%%%%%%%%%%%%%%%%%%%%%%%%%%%%%
The general crystal structure of a  monolayer TMDC (Fig. \ref{1}) consists of an hexagonal lattice with lattice parameter $a$ and a basis of three atoms: one  corresponding to a transition metal located at the lattice points and two  out of plane chalcogen atoms~\cite{kormanyos2015}. This structure (when viewed normally to the plane)  resembles that of graphene. The Brillouin zone (BZ) is hexagonal with two nonequivalent high symmetry points $K$ and $K'$ (see Fig.~\ref{1}).  One important departure from graphene is the lack of inversion symmetry, which results in a semiconductor gap whose magnitude depends on the particular kind of material. 
From hereon we will consider the case of WS$_2$ for the sake of concreteness, but our conclusions apply to the other members of the family as well.

%%%%%%%%%%%%%%%%%%%
\begin{figure}[t]
\begin{center}
 \includegraphics[width=0.8\columnwidth]{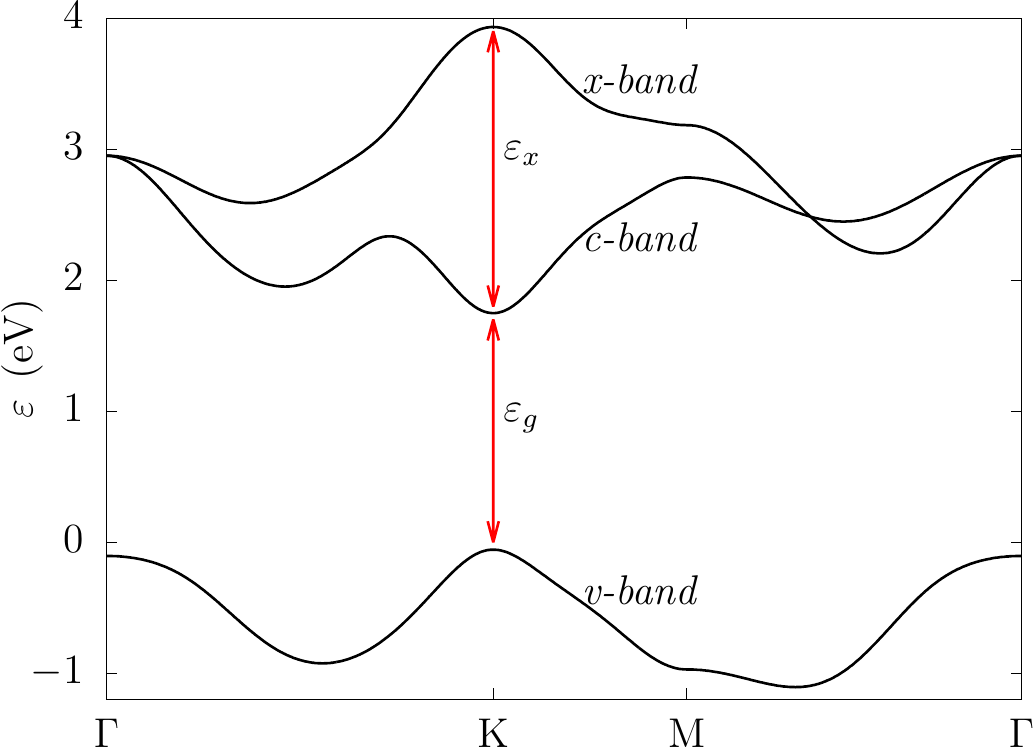}
 \caption{(Color online) Band structure for WS$_2$ along the path $\Gamma\rightarrow $K$ \rightarrow M \rightarrow \Gamma$ in the Brillouin zone, showing the direct semiconductor gap $\varepsilon_g$ at the $K$ point \label{banda}.}
\end{center}
\end{figure}
%%%%%%%%%%%%%%%%

Monolayer WS$_2$ has the $D_{3h}$ point group symmetry. Based on DFT calculations, the main contribution to the band structure of conduction and valence bands comes from W's $d$-orbitals, with a small contribution coming from S's $p$-orbitals. Different models for the band structure of TMDC have been put forward differing in the number of orbitals involved \cite{kormanyos2015,Dias2018,cappelluti2013}. Although initially it was common to treat this system in a two-band approximation~\cite{Xiao2012,Cao2012}, recently a three band model proved to account for a more complete description~\cite{Claassen2016}. Due to the marginal contribution from the S's $p$-orbitals it is possible to make a model using $d$-orbitals from W only. In addition,   due to the $z\rightarrow -z$ symmetry in the monolayer case, $d_{xz}$ and $d_{yz}$ orbitals (odd with respect to $z\rightarrow -z$ operation) are decoupled from $d_{xy}$, $d_{z^2}$ and $d_{x^2}$ orbitals, so that  a consistent model can be construct on the basis of the latter. Along these lines, and following Liu \textit{et al}~\cite{liu2013},  we use a three-band model composed of the $d_{xy}$, $d_{z^2}$ and $d_{x^2}$ W-orbitals with hoppings up to third nearest neighbors, necessary for a complete description over the whole BZ. Some details of this tight-binding model can be found in appendix A. Fig.~\ref{banda} shows the bands for monolayer WS$_2$ along the path $\Gamma$-$K$-M-$\Gamma$ in the BZ. We labeled the three bands as valence ($v$-band), conduction ($c$-band) and $x$-band. Notice that monolayer WS$_2$ is a direct gap semiconductor with a gap of $\varepsilon_g=1.81\,$eV at the $K$ point (and similarly for the $K'$ point). It will be useful for what follows  to also define the energy difference $\varepsilon_x=2.18\,$eV between the top of $x$-band and bottom of $c$-band.

%%%%%%%%%%%%%%%%%%%%%%%%%%%%%%%%%%%%%%%%%%%%%%%%%%%%%%%%%%%
\subsection{Equilibrium Edge States in Nanoribbons}\label{EqEdgeStates}
\begin{figure}[t]
\begin{center}
 \includegraphics[width=0.8\columnwidth]{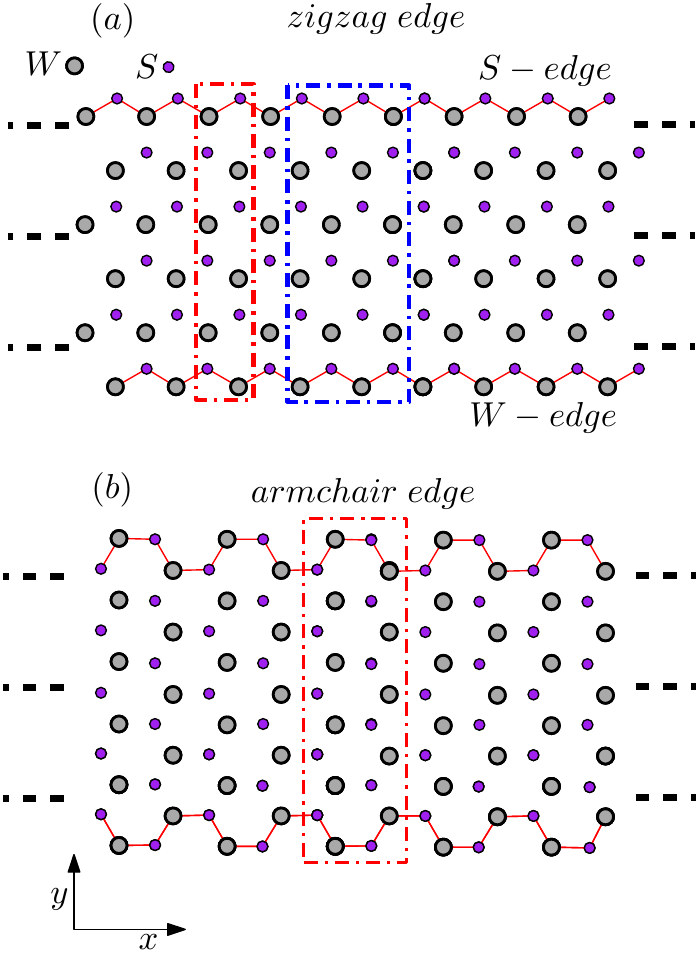}
 \caption{(Color online) Two kinds of edge termination for a WS$_2$ nano ribbon. (a) Zigzag edge where it is clear that both edges are different since they end in different types of atoms, W (bottom) or S (top). The red dotted square indicates the unit cell for density of state calculation while the blue one is the unit cell for calculations of conductance. (b) Armchair case. Here,  there is a reflection symmetry  through a plane normal to the layer and parallel to the edge \label{lattice}.}
\end{center}
\end{figure}

%%%%%%%%%%%%%%%%%%%%%%%%%%%%%%%%%%%%%%%%%%%%%%%%%%%%%%%%%%%
The WS$_2$ nanoribbons present themselves edge states with energy dispersion relations that depend on the kind of edge termination, namely zigzag or armchair \cite{farmanbar2016}. The zigzag case has a striking difference with that of graphene since in TMDC ribbons each edge is different: while one of them is composed of tungsten (W) atoms, the other is made of sulphur (S) atoms. We will denote them as W-edge and S-edge, respectively. This is clearly seen in Fig.~\ref{lattice}(a). As a result, edge states corresponding to different edges have different dispersion relations (see below), and will be treated separately.

Other important kind of edge is the armchair edge. One ribbon with this kind of termination is shown in Fig.~\ref{lattice}(b). This ribbon exhibits a symmetry of rotation of 180$^{\circ}$ around an in-plane axis parallel to the edge (equivalent to a reflection through a plane normal to the monolayer and parallel to the direction of the edge); this guarantees that both the local density of states and the dispersion relation be the same for both edges (see below). %Another form of ribbon with this edge can be constructed 

%%%%%%%%%%%%%%%%%%%%%%%%%
%%%%%%%%%%%%%%%%%%%%%%%%%

%%%%%%%%%%%%%%%%%%%%%
\begin{figure}[tpb]
\begin{center}
 \includegraphics[width=.9\columnwidth]{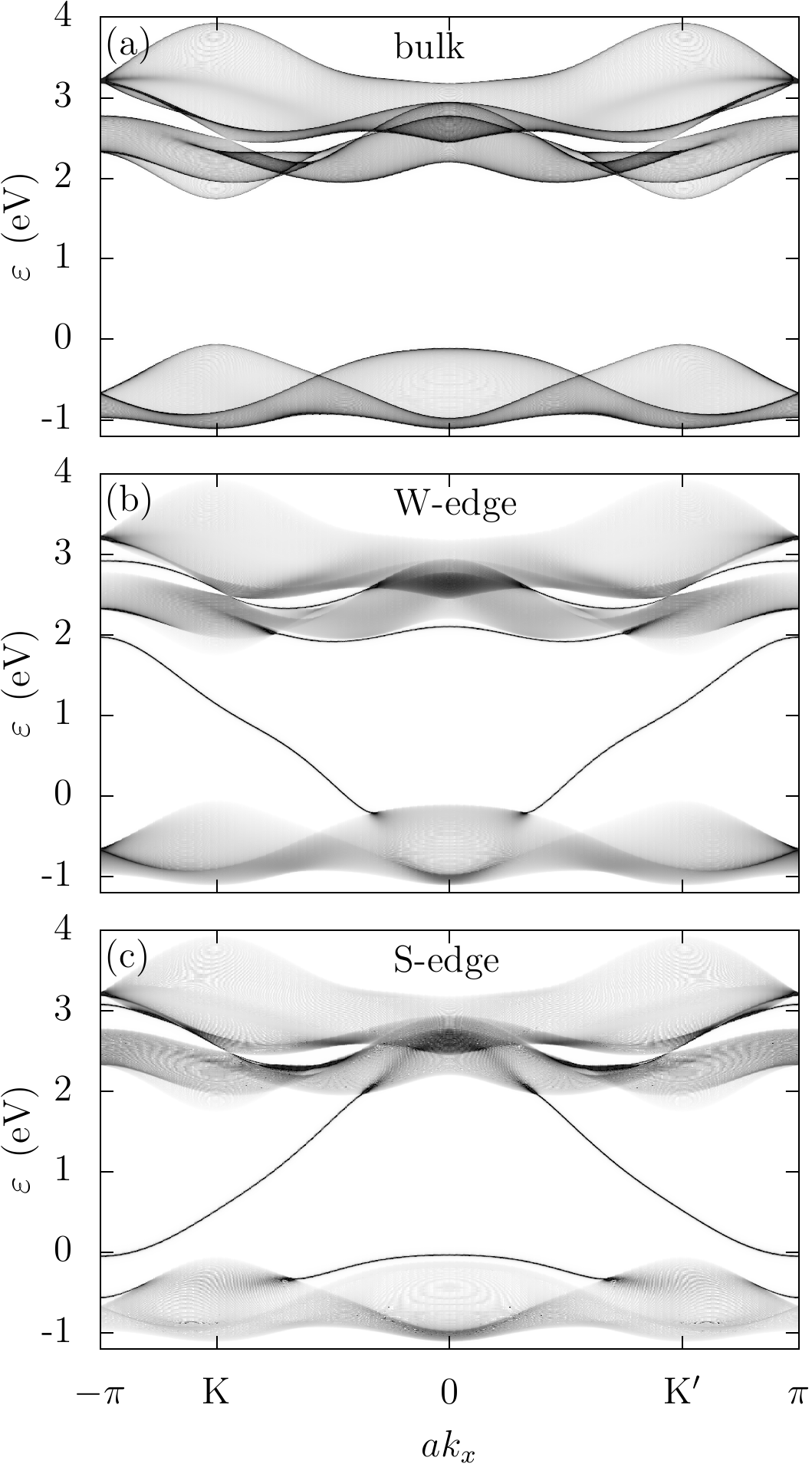}
 \caption{(Color online) Band structure ($k$-LSD) for a zigzag ribbon for bulk (a), the tungsten edge (b) and the sulphur edge (c). The grey scale indicates the local density of states. Note that each edge supports two edge states that cross the gap and travel in opposite directions. The energy dispersion of these edge states are different in both edges due to the difference in the lattice structure as seen in Fig. \ref{lattice} \label{zigzag}.}
\end{center}
\end{figure}
%%%%%%%%%%%%%%%%%%%%%

\begin{figure}[tpb]
\begin{center}    
 \includegraphics[width=0.9\columnwidth]{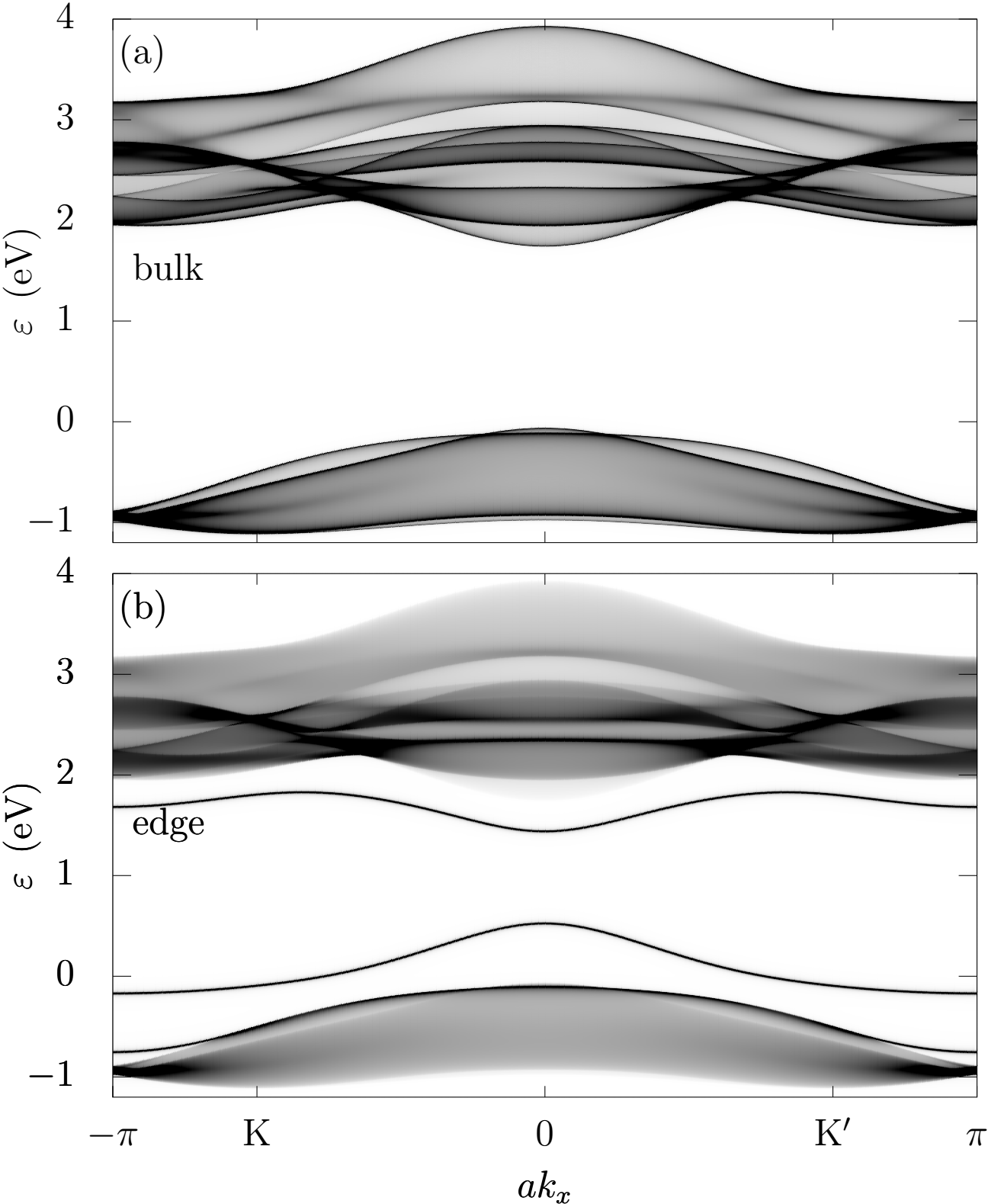}
 \caption{(Color online) Band structure ($k$-LSD) for a armchair ribbon for bulk (a) and one of the edges (b). In contrast with the zigzag ribbon, here both edges show the same  dispersion relations. Moreover, these states do not span the semiconductor gap and remain near both the conduction and valence band \label{armchair}.
 }
\end{center}       
\end{figure}

Figs. \ref{zigzag} and \ref{armchair} show the Local (in the tranverse site index) Spectral Density as a function of $k_x$ ($k$-LSD) for very wide zigzag and armchair ribbons, respectively~\cite{Usaj2014a}. This is done using a decimation method for the Green's function~\cite{Pastawski2001}. 
In the zigzag case there are metallic edge states spanning the entire semiconductor gap;  for each type of edge (W- or S-like) there are two edge states traveling with opposite velocities (positive or negative slop of the energy dispersion). In addition, there are also edge states near the top of the valence band for the S-edge and and near the bottom of the conduction band for the W-edge. Some extra edge states are also found in the energy region between the conduction and the $x$-band. 

For an armchair ribbon we have semiconducting edge states which do not close the gap (see Fig.~\ref{armchair}), and thus leave the ribbon with an effective smaller gap. Due to the symmetry mentioned before, we only show the bands for one of the edges only.

%%%%%%%%%%%%%%%%%%%%%
%%%%%%%%%%%%%%%%%%%%%

%%%%%%%%%%%%%%%%%%%%%%%%%%%%%%%%%%%%%%%%%%%%%%%%%
\section{Floquet States \label{floquetstates}}
%%%%%%%%%%%%%%%%%%%%%%%%%%%%%%%%%%%%%%%%%%%%%%%%%
In this Section we describe how circularly polarized radiation can induce topological states on WS$_2$, hence leading to a Floquet topological insulator~\cite{Oka2009,Kitagawa2010,Lindner2011,Rudner2013}.
%%%%%%%%%%%%%%%%%%%%%%%%%%%%%%%%%%%%%%%%%%%%%%%%%%%%%%%%%%
\subsection{Floquet Formalism \label{floquetformalism}} %%%%
%%%%%%%%%%%%%%%%%%%%%%%%%%%%%%%%%%%%%%%%%%%%%%%%%%%%%%%%%%
When the Hamiltonian  does depend on time explicitly, the energy of the system is no longer a conserved quantity and the usual approach of diagonalizing the Hamiltonian is no longer useful. Yet, for the special cases where the Hamiltonian is periodic in time one can apply the Floquet theory to reduce the calculation once again to an eigenvalue problem. Just as a brief introduction we will present here the basic features of this method.

Floquet theory \cite{Shirley1965,Sambe1973,Grifoni1998a,Kohler2005} is a suitable approach for problems involving periodic time dependent  Hamiltonians $H(t)=H(t+2\pi/\Omega)$, which can be written as a Fourier series in time $H(t)=\sum_m H^{(m)}\,e^{i\,m\Omega t}$. The solutions of the time dependent Schr{\"o}dinger equation $i\hbar\,\partial_t|\Psi\rangle=H(t)|\Psi\rangle$ can be written as $|\Psi(t)\rangle=e^{-i\,\varepsilon t/\hbar}\,|\Phi(t)\rangle$, with $|\Phi(t)\rangle$ periodic in time with the same period as $H(t)$. The quantity $\varepsilon$ is called the quasienergy. With this {\it ansatz} $|\Phi(t)\rangle$ satisfies the so-called Floquet equation
%%%%%%%%%%%%%%%%%%%
\begin{equation}
(H(t)-i\hbar\,\partial_t)|\Phi(t)\rangle=\varepsilon|\Phi(t)\rangle\,.
\end{equation}
%%%%%%%%%%%%%%%%%%%
 Since $|\Phi(t)\rangle$ is periodic in time we can treat this time dependent problem as a time independent one by extending the Hilbert space as a product $\cal{R}\otimes\cal{T}$ of the usual time independent space $\cal{R}$ (of the static system) and the space $\cal{T}$ of functions periodic in time with period $T=2\pi/\Omega$. $\cal{T}$ space can be spanned by the $|m\rangle$ basis functions that satisfy
%%%%%%%%%%%%%%%%%%%%%%  
\begin{equation}\label{floq}
\langle t|m \rangle=\frac{1}{\sqrt{2\pi}}e^{i m\Omega t},\;m=0,\pm 1,\pm 2,\cdots\,,
\end{equation} 
%%%%%%%%%%%%%%%%%%%%%%
when projected onto the time basis. 
The final set of eigenfunctions are then written as $|\chi,m\rangle$, where $\chi$ refers to the orbitals $d_{z^2}$, $d_{xy}$ or $d_{x^2}$, and $m$ to the {\it Floquet replica} or Floquet subspace and runs according to Eq.~\eqref{floq}, $|\Phi(t)\rangle=\sum_{\chi,m}  e^{i m\Omega t} c_{\chi,m}|\chi,m\rangle$. It must be said that when looking at the $K$-points, it would be more useful to resort to the basis given by Eqs.~\eqref{b1} and~\eqref{b2}, which diagonalize the static Hamiltonian at these points and are written as $|v\rangle$, $|c\rangle$ and $|x\rangle$, refering to the valence, conduction and $x$-band respectively (see Appendix B for details).   By expanding the time dependent Hamiltonian in a Fourier series we can construct the Floquet Hamiltonian $H_F$ (cf. Eq.~\eqref{HF}). We use the $3\times 3$ tight-binding Hamiltonian Eq.~\eqref{hamilton}. To apply the Floquet method in the present case we make use of the well known Peierls substitution. Namely, we make the replacement
%%%%%%%%%%%%%%%%%%%%%%%%%%%
\begin{equation}\label{11}
\hbar\bm{ k} \rightarrow \hbar\bm{ k} + \frac{e}{c}\,\bm{ A}(t)\,.
\end{equation}
%%%%%%%%%%%%%%%%%%%%
Since we are mainly interested on the effects of circularly polarized light we take the vector potential to be $A_0\,(\cos(\Omega t)\, \hat{\bm{x}}+\sin(\Omega t+\varphi)\, \hat{\bm{y}})$. The $\varphi$ argument can take the values $0$ and $\pi$, indicating anticlockwise and clockwise polarization, respectively. With this choice the Peierls substitution gives place to terms in the Hamiltonian of the general form $\cos(\alpha k_x+\beta k_y+\alpha A_0\,\sin\Omega t+\beta A_0\,\cos\Omega t)$ and $\sin(\alpha k_x+\beta k_y+\alpha A_0\,\sin\Omega t+\beta A_0\,\cos\Omega t)$, $\alpha$ and $\beta$ being real constants.  These terms can readily be Fourier expanded by using the well known Jacobi-Anger identity \cite{watson}
%%%%%%%%%%%%%%%%%%%%%%%%%%%%%%%%%%
\begin{equation}
   \label{jacobi}
   e^{i z\sin\theta}  = \sum_n\, \,J_n(z)\,e^{in\theta}\,,
\end{equation}
%%%%%%%%%%%%%%%%%%%%%%%%%%%%%%%%%%
where $J_n(z)$ is the Bessel function of the first kind with integer order $n$. In expanding these expressions we will encounter double summations of Bessel functions that can be simplified (and thus reduce the computational effort) with the following property of Bessel functions \cite{watson}
%%%%%%%%%%%%%%%%%%%%%%%%%%%%%%%%
\begin{equation}\label{suma}
\sum_m J_{n+m}(\alpha)J_m(\beta)\,e^{-im\Omega }=e^{-in\Psi}J_n(\Gamma)\,,
\end{equation}
%%%%%%%%%%%%%%%%%%%%%%%%%%%%%%%
with $\Gamma=\sqrt{\alpha^2+\beta^2-2\alpha\beta\cos\Omega}$ and $\Psi$ given by $\tan\Psi=\beta\sin\Omega/(\alpha-\beta\cos\Omega)$. Hence, the Floquet Hamiltonian $H_F$ can be written (in the $|\chi,m\rangle$ basis) in the usual block matrix form:
%%%%%%%%%%%%%%%%%%%%%%%%%%%% 
\begin{equation}\label{HF}
H_F=\left[
\begin{array}{ccccc}
\ddots & \vdots          & \vdots  & \vdots          & \reflectbox{$\ddots$} \\
\dots  & H^{(0)}+\hbar\Omega & H^{(1)}     & H^{(2)}             & \dots \\
\dots  & H^{(-1)}          & H^{(0)}     & H^{(1)}             & \dots \\
\dots  & H^{(-2)}          & H^{(-1)}  & H^{(0)}-\hbar\Omega & \dots \\
\reflectbox{$\ddots$} & \vdots          & \vdots  & \vdots          & \ddots
\end{array}
\right]\,.
\end{equation}
%%%%%%%%%%%%%%%%%%%%%%%%%%%%
Here $H^{(m)}$ is the $m$-th Fourier component of $H(t)$.

%%%%%%%%%%%%%%%%%%%%%%%%%%%%%%%%%%%%%%%%%%%%%%%%%%%%%%%%%%%%%%%%%%%
%%%%%%%%%%%%%%%%%%%%%%%%%%%%%%%%%%%%%%%%%%%%%%%%%%%%%%%%%%%%%%%%%%%
\subsection{Floquet Bulk Bands\label{FloquetBulkBands}}%%%%%%%%%%%%
%%%%%%%%%%%%%%%%%%%%%%%%%%%%%%%%%%%%%%%%%%%%%%%%%%%%%%%%%%%%%%%%%%%
%%%%%%%%%%%%%%%%%%%%%%%%%%%%%%%
\begin{figure}[tpb]
\includegraphics[width=0.9\columnwidth]{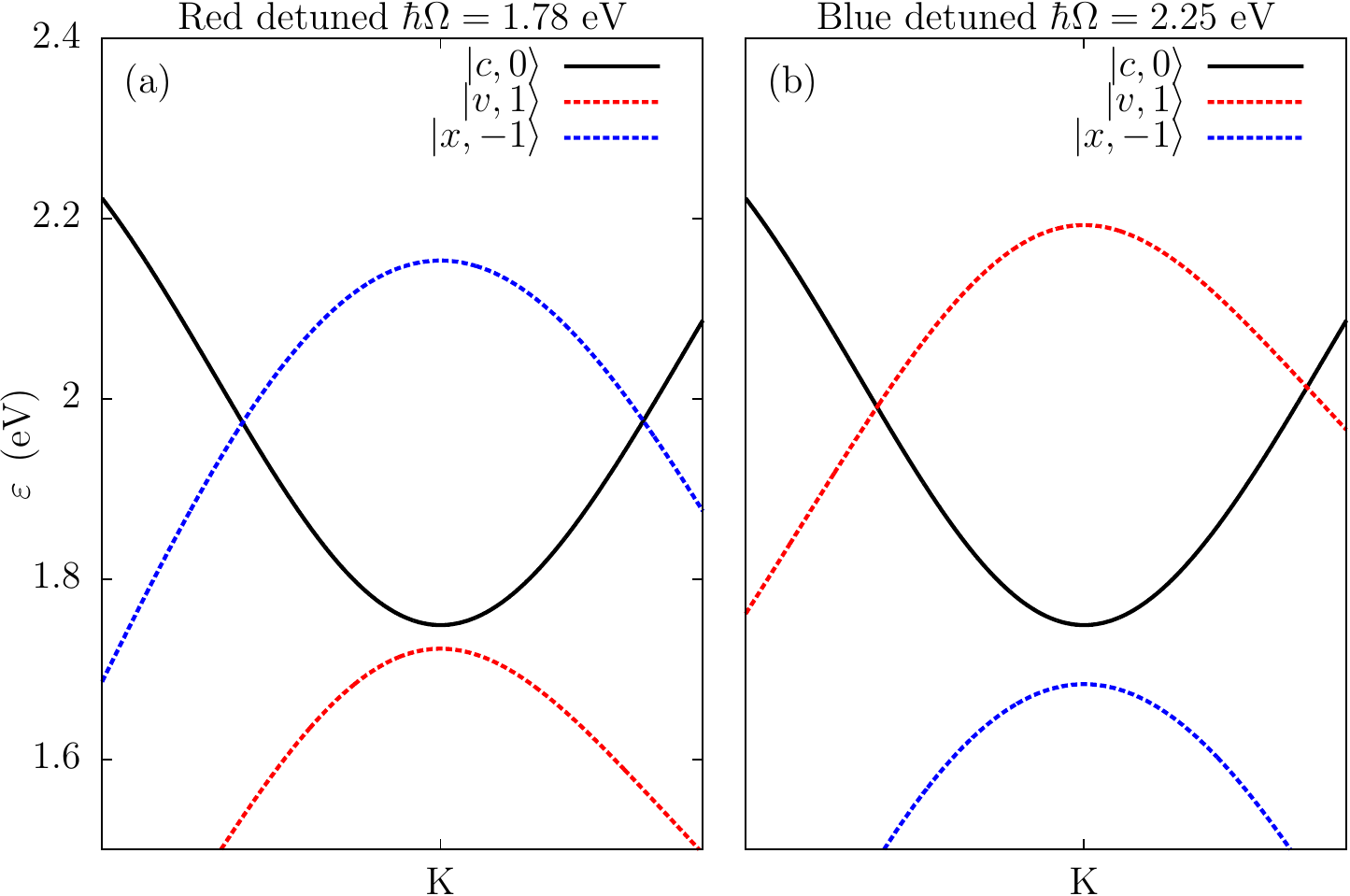}
\caption{(Color online) The two regimes of photon energy $\hbar\Omega$: (a) Red detuned with $\hbar\Omega< 1/2(\varepsilon_x(\bm{K})-\varepsilon_v(\bm{K}))$, which brings into resonance the conduction and $x$ bands, while leaving the valence bands below the conduction one, that is ensured by the extra condition $\varepsilon_v(\bm{K})+\hbar\Omega<\varepsilon_c(\bm{K})$. (b) The blue detuned regimen is characterised  by $\hbar\Omega>1/2(\varepsilon_x(\bm{K})-\varepsilon_v(\bm{K}))$, and this gives place to a resonance between conduction and valence bands. The additional condition $\varepsilon_x(\bm{K})-\hbar\Omega<\varepsilon_c(\bm{K})$  pushes the $x$ band bellow the conduction band and off resonance \label{19}.}
\end{figure}
%%%%%%%%%%%%%%%%%%%%%%%%%%5
Following Ref.~\cite{Claassen2016}, when studying the Floquet bands in TMDCs it is convenient to define two regimes for the possible values of $\hbar\Omega$: the blue and red detuned regimes. 
These two regimes are defined by the positions of the Floquet replicas $|v,1\rangle$ and $|x,-1\rangle$ with respect to the conduction band $|c,0\rangle$. If $\varepsilon_v(\bm{ k})$, $\varepsilon_x(\bm{ k})$ and $\varepsilon_c(\bm{ k})$ are the corresponding dispersion relations for our three band model, then the red-detuned regime corresponds to the case $\varepsilon_x(\bm{ K})-\hbar\Omega > \varepsilon_v(\bm{ K})+\hbar\Omega$, whereas the blue detuned is defined by $\varepsilon_x(\bm{K})-\hbar\Omega < \varepsilon_v(\bm{ K})+\hbar\Omega$. The transition between both regimes occurs at $\hbar\Omega_c=(\varepsilon_x(\bm{K})-\varepsilon_v(\bm{K}))/2$.  
Choosing appropriately the value of $\hbar\Omega$ we can make, for instance, the conduction band to go into resonance with only one of the Floquet replicas, $|x,-1\rangle$ or $|v,1\rangle$, as it is shown in Fig. \ref{19}. For the specific case of WS$_2$ ($\varepsilon_g=1.81$ eV and $\varepsilon_x=2.18$ eV) we have that $\hbar\Omega_c=(\varepsilon_x+\varepsilon_g)/2\sim2$\,eV.

\begin{figure}[tpb]
\includegraphics[width=0.95\columnwidth]{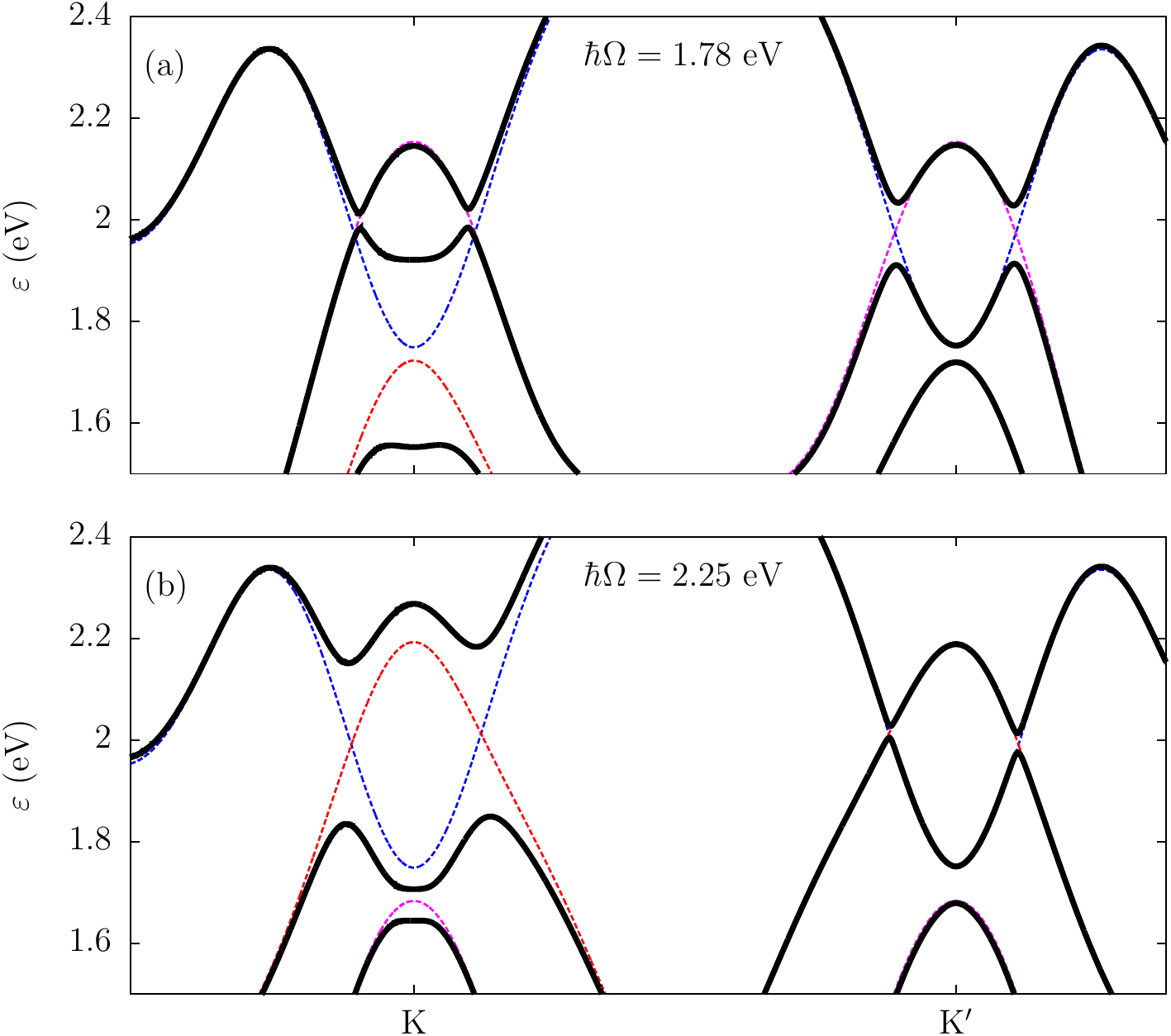}
\caption{ (Color online) Floquet bulk states in two different regimes and with $s=0.05$ in both cases. In any case we use five Floquet replicas with $-2 \leq m \leq 2$. (a) Red detuned with $\hbar\Omega=1.78\,$eV showing the optical Stark shift between states $|c,0\rangle$ and $|v,+1\rangle$. This shift is large when compared with the gap opening in the crossing between $|c,0\rangle$ and $|x,-1\rangle$.  (b) Blue detuned regimen with $\hbar\Omega=2.25\,$eV. In both cases it is clear that only a certain $K$ point is sensitive to the laser field, depending on the direction of rotation of the vector potential $\bm{ A}(t)$ \label{12}.}
\end{figure}

Optical selection rules tell (see Appendix B) us that an anticlockwise vector field $\bm{ A}(t)=A_0\,(\cos\Omega t\,\hat{\bm{x}}+\sin\Omega t\,\hat{\bm{y}})$ can only couple the transitions $|v,1\rangle \rightarrow |c,0\rangle \rightarrow |x,-1\rangle$  at the $K$ point, whereas at $K'$ those transitions are forbidden.   
For the red-detuned regime (Fig.~\ref{12}(a))  there is a sizable band repulsion (optical Stark shift) between $|c,0\rangle$ and $|v,1\rangle$  at exactly the $K$ point while leaving  $|x,-1\rangle$ almost unchanged. Selection rules predict this upshot for this particular choice of $\bm{ A}(t)$, see appendix \ref{StarkShift}, although the negligible shift in $|x,-1\rangle$ needs to be accounted for by using perturbation theory. At $K'$ the Floquet replicas are not modified (the same is valid in the blue detuned case, see Fig.~\ref{12}(b)).  In a neighborhood of $K$ and $K'$ the optical selection rules no longer hold exactly but approximately. In the vicinity of both of these $K$ points, where the conduction and the replica of the $x$ band cross each other, small gaps develop. We refer to them as dynamical gaps. They host chiral edge states as we will discuss below.
%%%%%%%%%%%%%%%%%%%%%%%%%%%%%%%
%%%%%%%%%%%%%%%%%%%%%%%%%%%%%%%%%%

%%%%%%%%%%%%%%%%%%%%%%%%%%%%%%%%%%
\subsection{Floquet Edge States\label{FloquetEdgeStates}}
%%%%%%%%%%%%%%%%%%%%%%%%%%%%%%%%%%%%%%%%%%%%%%%%%
In the previous section we have discussed the gaps opening in the bulk Floquet bands, whose properties can be deduced from the analysis of the tight binding Hamiltonian in $\bm{k}$-space presented in Eq.~\eqref{hamilton} after the Peierls substitution is done. We will now explore the question of whether we can find edge states of  topological nature inside those gaps, in a similar fa\-shion as obtained in monolayer graphene \cite{Perez-Piskunow2014,Usaj2014a,Perez-Piskunow2015}.

In order to study a finite width ribbon we need to go back to the real space lattice tight binding Hamiltonian. The details of its construction are quite standard. In this case, the time dependence induced by the radiation field is introduced by changing the hopping matrix elements according to the following form of Peierls substitution:
%%%%%%%%%%%%%%%%%%%%
\begin{equation}\label{exponente}
t_{ij}  \rightarrow  t_{ij}\,e^{ie/\hbar c\,\int_{\bm{r}_i}^{\bm{r}_j} \bm{ A}(t)\cdot d\bm{\ell}}=t_{ij}\,e^{ie/\hbar c\,\bm{ A(t)}\cdot \bm{ R}_{ij}},
\end{equation}
%%%%%%%%%%%%%%%%%%
which is equivalent to that given in Eq.~\eqref{11}. The last equality holds since $\bm{A}(t)$ is homogeneous in space, $\bm{ R_{ij}}={\bm{r}_j}-{\bm{r}_i}$ being the lattice vector between  sites $i$ and $j$. 
%In the following we will take $\bm{ A}(t)=A_0\,(\cos\Omega t\,\hat{\bm{x}}+\sin\Omega t\,\hat{\bm{y}})$ (anticlockwise). 
It must be emphasized that the nomenclature $i$ and $j$ refers, more precisely, to both the $d$-orbital ($d_{z^2}$, $d_{xy}$ or $d_{x^2}$) and the lattice point. The Fourier expansion of the Hamiltonian depends ultimately  on the expansion of the exponential in Eq.~\eqref{exponente}. This can be done using the same properties of the Bessel functions given in Eqs.~\eqref{jacobi} and (\ref{suma}). For the calculations of Floquet spectra we will use the decimation method for the Green functions (see Refs. [\onlinecite{Pastawski2001}] and [\onlinecite{farmanbar2016}] for details).

We performed our calculations with photon energies of the order of the gap size ($\sim2$ eV), in order to couple the replicas of the $x$ and $v$ bands with the conduction band to first order,  and with small intensities, which enters through the dimensionless parameter $s=eA_0a/2\hbar c$. We allowed up to second order photon processes, which means that we used Floquet replicas from $m=-2$ through $m=+2$. In order to prove the sufficiency of these figures, we performed subsidiary calculations with more replicas and verified that we get an acceptable convergence with just five of them.
%%%%%%%%%%%%%%%%%%%%%%%%%%%
\begin{figure}[tpb]
\includegraphics[width=0.95\columnwidth]{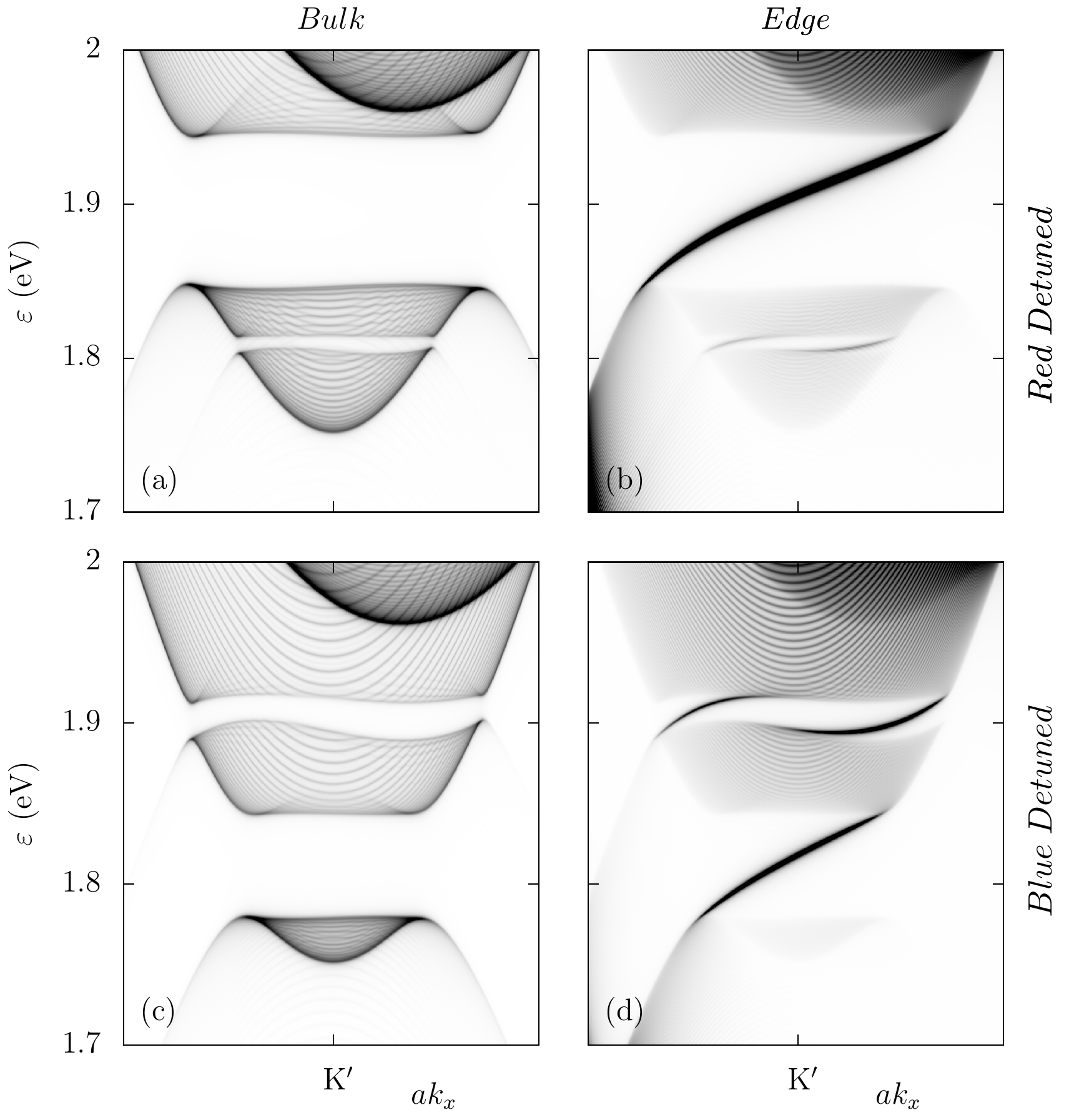}
\caption{Topological transition of the $0$-Floquet spectrum when going from red detuned ((a) and (b)) to blue detuned ((c) and (d)) in a zigzag ribbon. The bands are shown in a vecinity of the $K'$ point and are  projected over the subspace $m=0$ (in the range of quasi energies corresponding to the equilibrium conduction band). Sub-figures (a) and (b) correspond to $\hbar\Omega=1.92\,$eV and $s=0.05$ (red detuned), and in this case the uppermost gap results from the coupling between replicas $|c,0\rangle$ and $|x,-1\rangle$; the parameters for (c) and (d) are $\hbar\Omega=2.07\,$eV and $s=0.05$ (blue detuned) and now the coupling of $|c,0\rangle$ and $|v,1\rangle$ generates the uppermost gap. (a) and (c) are bulk bands and (b) and (d) are edge bands (one edge only for clarity). In any case the coupling between  $|x,-1\rangle$ and $|c,0\rangle$ gives place to one chiral edge state, whereas coupling $|c,0\rangle$ and $|v,+1\rangle$ gives two edge states. The topological invariants will be calculated in any case for the uppermost of these gaps \label{14}.}
\end{figure}
%%%%%%%%%%%%%%%%%%%
As in graphene, irradiated WS$_2$ ribbons harbor edge states inside the dynamical gaps, though in the present case the edge states appear  only near the $K'$ point for the chosen polarization (anticlockwise). This is more evident in the zigzag termination where the two cones do not overlap. Here, we concentrate in the states formed by the crossing of replicas $|c,0\rangle$ and $|x,-1\rangle$ on the one hand (red detuned), and $|c,0\rangle$ and $|v,1\rangle$ on the other hand (blue detuned). Fig.~\ref{14} shows the Floquet $k$-LSD weighted over the $m=0$ replica ($0$-Floquet band from now on) for bulk and one edge for the red and blue detuned cases. It is apparent from these figures that the coupling of the $|x,-1\rangle$ and $|c,0\rangle$ bands  gives place to one chiral edge state, whereas the coupling of $|c,0\rangle$ and $|v,+1\rangle$ gives two chiral edge states. If we consider the uppermost gap in both cases, it is clear that there is a topological transition that changes the number of states appearing inside such a gap and that this transition occurs when passing from the red detuned to the blue detuned regime~\citep{Claassen2016}. Similar considerations can be drawn from other gaps but is in this one where we can see the effects in their more pristine form. 
%%%%%%%%%%%%%%%%%%%%%%%%%%%%%%%%%%%%%%%%%%%%%%%%
\subsection{Topological Characterization}%%%%%%%
%%%%%%%%%%%%%%%%%%%%%%%%%%%%%%%%%%%%%%%%%%%%%%%
%%%%%%%%%%%%%%%%%%%%%%%%%
\begin{figure}[t]
\includegraphics[width=0.95\columnwidth]{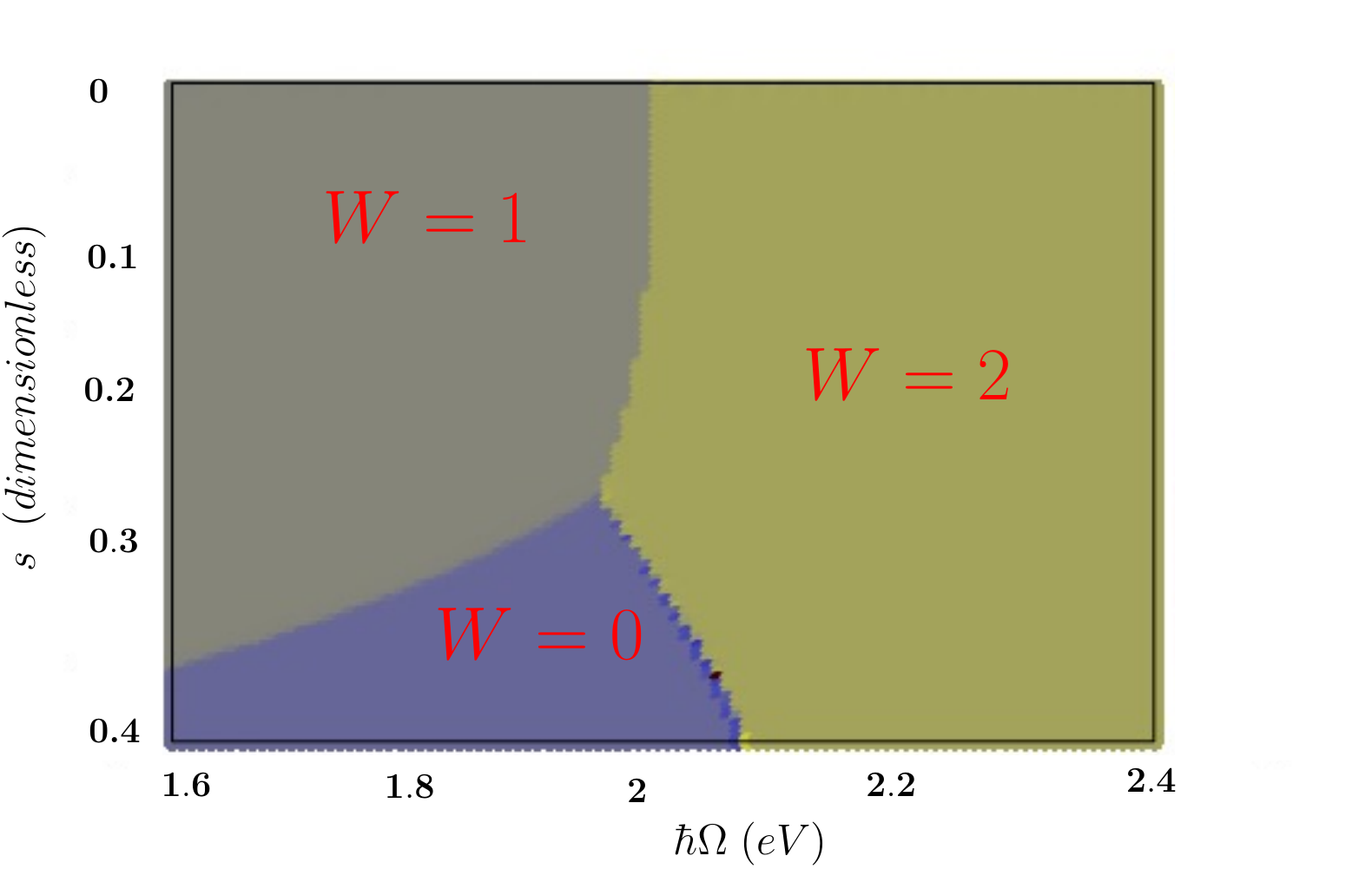}
\caption{(Color online) Topological phase diagram showing the transition from one to two chiral edge states ($s=eA_0a/2\hbar c$). The two regions correspond to the red and blue detuned regimes before mentioned \label{15}.}
\end{figure}
%%%%%%%%%%%%%%%%%%%%%%%%%%%%%%%%%%%%%%
The topological nature of the Floquet edge states can be deduced from the bulk band structure by looking at the  Chern numbers $C_n$ associated to each one of the Floquet bands, labeled here with $n$. Their calculation involves an integration over the whole BZ of the Berry curvature $\bm{\Gamma}_n(\bm{k})$
%%%%%%%%%%%%%%%%%%%%%%%%%%%%%%%%%%%%%%
\begin{eqnarray}\label{chern}
C_n &=&\frac{1}{2\pi}\int_{\mathrm{BZ}} dk_xdk_y\,\bm{\Gamma}_n({\bm k})\cdot\hat{\bm{z}}\,,\\
\nonumber
\bm{\Gamma}_n({\bm k}) &=&\sum_{m\neq n} \mathrm{ Im} \frac{\langle u_n|\bm{\nabla}_{\bm k} H_\mathrm{F}|u_m \rangle \times
\langle u_m|\bm{\nabla}_{\bm k} H_\mathrm{F}|u_n \rangle}{(\varepsilon_n-\varepsilon_m)^2}\,.
\end{eqnarray}
%%%%%%%%%%%%%%%%%%%%%%%%%%%%%%%%%%%%%%
As it is well known and has been extensively discussed in the literature \cite{Rudner2013,Nathan2015,Carpentier2015,Perez-Piskunow2015}, in contrast with static systems, the bulk-edge correspondence and its relation to the Chern numbers is very subtle when describing time dependent problems. As before, we construct a truncated Hamiltonian keeping the replicas ${-M\leqslant m\leqslant M}$, where $M$ is a large positive integer. The resulting  matrix  can be interpreted as the Hamiltonian  of a static system. With this reduced Floquet Hamiltonian we can calculate the net number of chiral edge states (those traveling in one direction minus those traveling opposite) in a given gap as a summation of the $C_n$ of all the bands below it:
%%%%%%%%%%%%%%%%%%%%%%%%%%
\begin{equation}
   W_n=\sum_{m\leq n} C_m.
\end{equation}
%%%%%%%%%%%%%%
Clearly, not all the Floquet  gaps are well defined so we will concentrate in the gap where we have seen the transition depicted in Fig.~\ref{14}. In Fig.~\ref{15} we show the winding number $W_n=W$ for the uppermost gap in Fig. \ref{14} as a function of the photon energy $\hbar\Omega$ and the dimensionless parameter $s$. 
In the range of values plotted, we clearly see three regions with different values of $W$. For small values of $s$ ($\lesssim$ 0.3),  a transition from $W=1$ to $W=2$ occurs near $\hbar\Omega=2\,$eV, with decreasing value as $s$ increases. This transition is in agreement with the $0$-Floquet bands shown in Fig.~\ref{14}. As $s$ increases above $0.3$, the transition line reaches a triple point where a region with $W=0$ appears. This gives place to two new transition lines separating the region $W=0$ from those with $W=1$ and $W=2$. 

%%%%%%%%%%%%%%%%%%%%%%%%%%%%%%%%%%%%%%%%%%%%%%%%%%%%%%%%%%%%%
\subsection{Effect on the Equilibrium Edge States\label{effectonequilibrium}}%%%%%%%%%%%
%%%%%%%%%%%%%%%%%%%%%%%%%%%%%%%%%%%%%%%%%%%%%%%%%%%%%%%%%%%%%
As shown in Figs.~\ref{zigzag} and~\ref{armchair}, WS$_2$ ribbons also support, in the absence of radiation, a set of non topological edge states. In the zigzag case they span the entire band gap, while for the armchair termination they do not. We will now analyze how these edge states are affected by the  laser field. 

%%%%%%%%%%%%%%%%%%%%%%%%%
\begin{figure}[t]
\includegraphics[width=0.8\columnwidth]{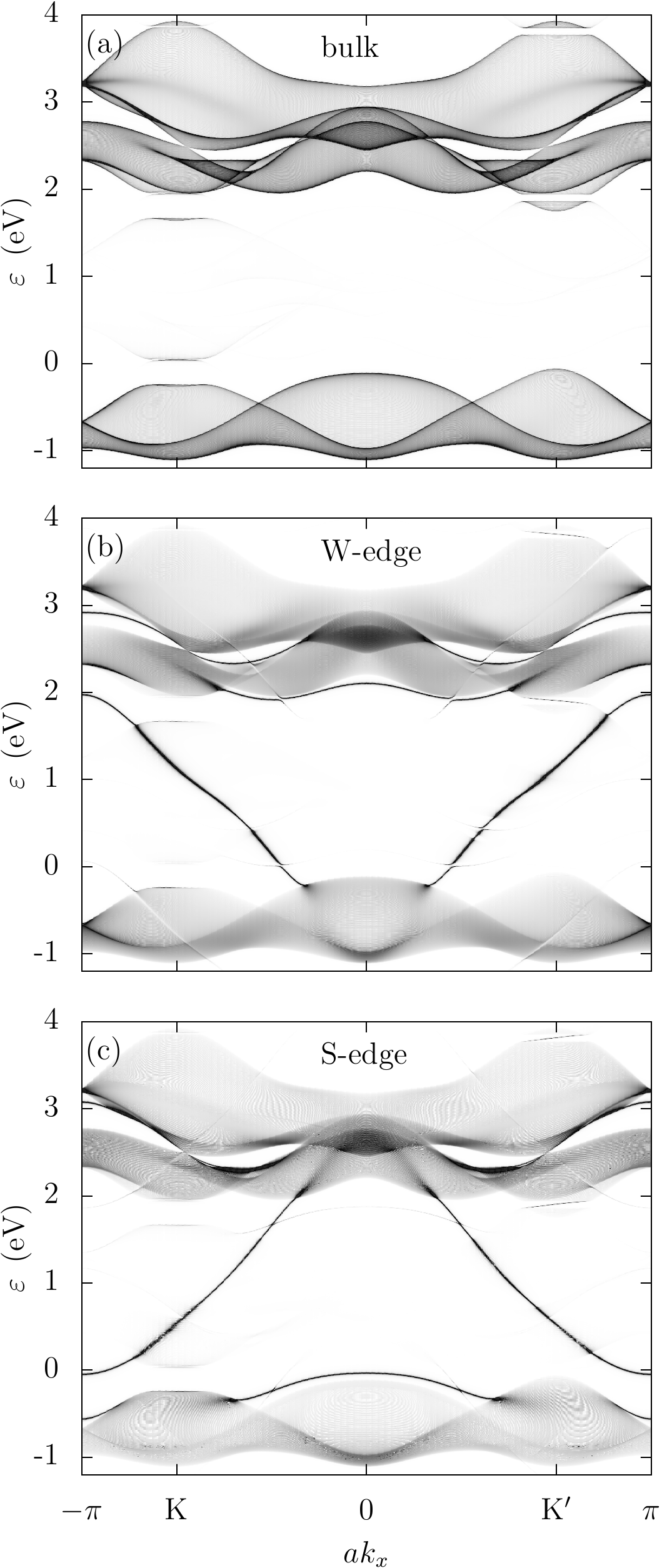}
\caption{$0$-Floquet bands (bulk (a) and edges (b and c)) of a wide ribbon (2050 tungsten atoms wide) in the red detuned regime for $\hbar\Omega=1.91\,$eV and $s=0.04$. Notice the appearance of chiral edge states inside the dynamical gap formed in the $K'$ valley. These bands have the property that making $k_x \rightarrow -k_x$ is equivalent to change the direction of rotation of the vector field $\bm{A}(t)$. It is worth  pointing out the emergence of gaps on the energy dispersion of the equilibrium W-edge states due to the resonant coupling with other edge states in the $c$- and $x$-bands. \label{irradiated-edges}}
\end{figure}
%%%%%%%%%%%%%%%%%%%%%%%%
%%%%%%%%%%%%%%%%%%%%%%%
\begin{figure}[t]
\includegraphics[width=0.8\columnwidth]{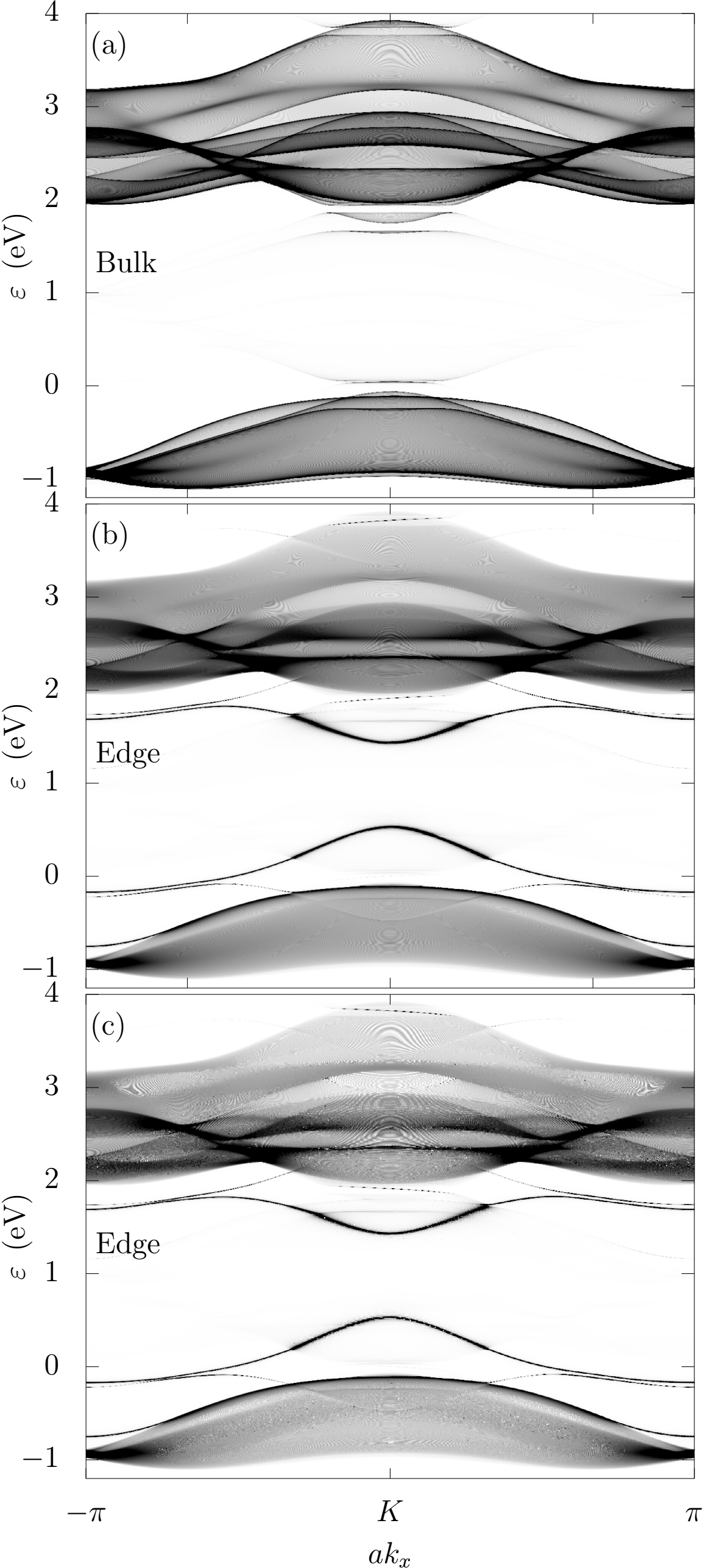}
\caption{The same as Fig.~\ref{irradiated-edges} but for the armchair edge \label{irradiated-edges-armchair}.}
\end{figure}
%%%%%%%%%%%%%%%%%%%%%%%

The results for the $0$-Floquet bands for a zigzag termination are shown in Fig.~\ref{irradiated-edges}. There are several interesting features to point out: (i) small gaps develop at points in $\bm{k}$-space where there is a resonance between the W-edge states inside the gap and those near the bottom of the $c$- and $x$-bands. The same situation occurs in the  the S-edge, where the replica of the narrow edge states near the valence band couples with the edge states near the conduction band, although this coupling is less favorable than in the W-edge and in order to make it apparent we require a higher intensity. This can be explained by looking at the orbital's character of the edge states: in the S-edge the almost flat edge states near the top of the valence band have their largest weight on $d_{z^2}$ orbitals, while those spanning the gap are mainly of $d_{xy}$ character. The coupling between these orbitals is given by $t_1\sim -0.1$ eV, while, as a comparison, in the W-edge the relevant hopping is $t_2\sim 0.6$ eV; (ii) the coupling between the equilibrium edge states and the replicas of the bulk bands leads also to a \textit{broadening} of the $k$-LSD of these edge states, that is a loss of the projected spectral weight, which in turn depends on the order of the photon transition (Floquet replica) involved; (iii) the latter effect is also different for the W- and S-edge states depending on the quasienergy range; iv) both the gaps and the broadening are selective: for a given polarization of the laser field, say anticlockwise, the equilibrium edge states traveling in one direction are much more affected than those that travel in the  opposite direction. The situation is, of course, inverted when the polarization is changed to be clockwise. 
Similar effects occur for the armchair termination (see Fig.~\ref{irradiated-edges-armchair}).

%%%%%%%%%%%%%%%%%%%%%%
\begin{figure}[t]
\includegraphics[width=0.95\columnwidth]{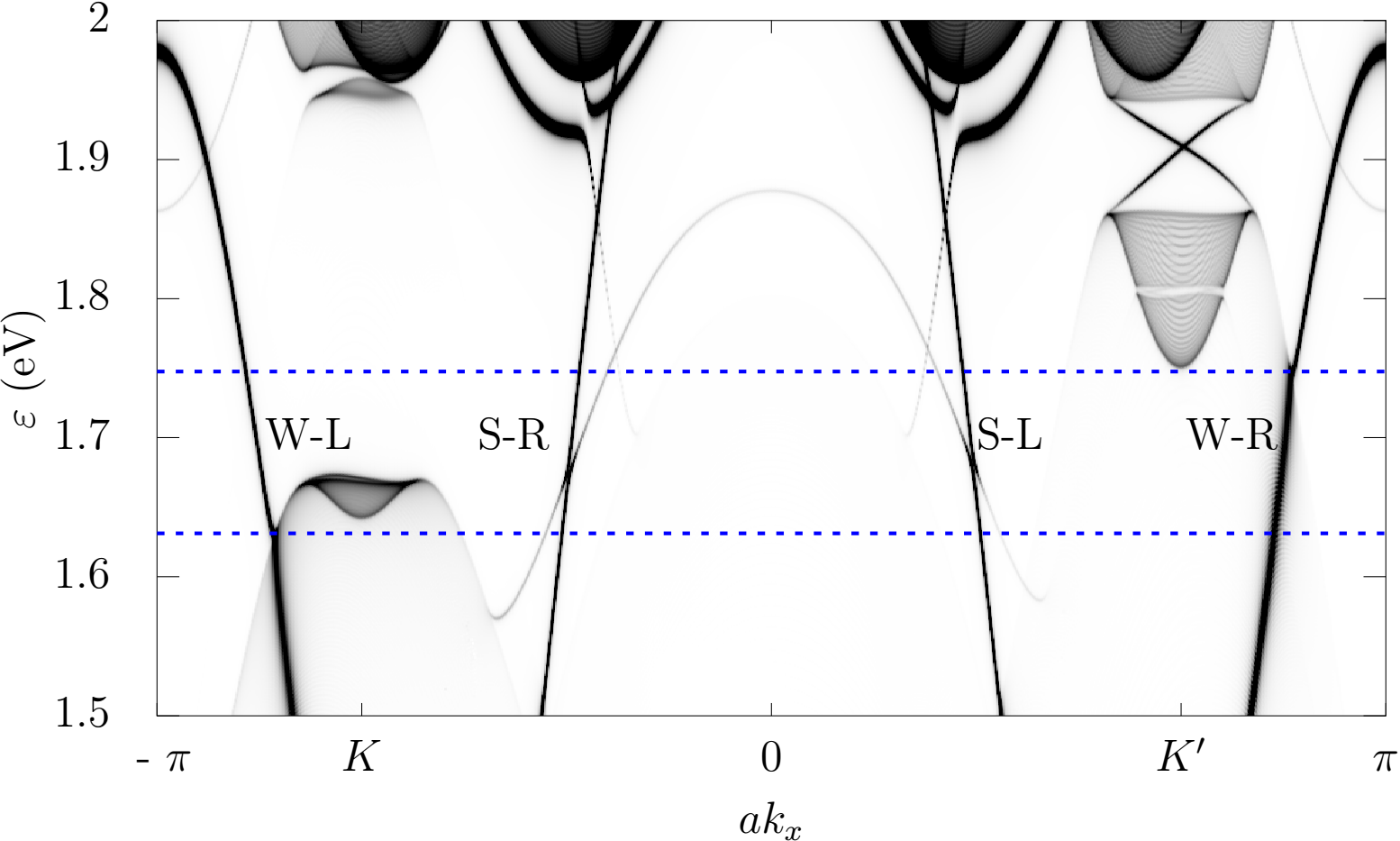}
\caption{Zoom in of the $0$-Floquet bands (adding the bulk and edges) for the ribbon of the previous figure, showing the selective {\it switching-off} of the equilibrium edge states (see text). \label{zoomFloquet}.} 
\end{figure}
%%%%%%%%%%%%%%%%%%%%%%

To better appreciate the effect of the broadening of the edge states, we plot in Fig.~\ref{zoomFloquet} a zoom of the $0$-Floquet bands (adding the bulk and both edges) in the quasienergy range ${1.5\,\mbox{eV}\leqslant \varepsilon \leqslant 2\,\mbox{eV}}$ and in the same red detuned regime as in Fig.~\ref{irradiated-edges}. The edge states in the S-edge are labelled as S-L and S-R, according to whether they travel to the left (S-L) or to the right (S-R). Similarly for the W-edge. In particular, we look at a given quasienergy region (delimited by the blue dashed horizontal lines) where the effect is clearer: the right-moving W-R state  is suppressed due to the coupling with the $m=1$ replica of the $v$-band. On the contrary, near the $K$ point, this $|v,1\rangle$ replica is pushed down due to the Stark effect and thus does not couple with the left moving W-L edge state, leaving it unaltered in that quasienergy range. On the other hand, both states S-L and  S-R remain untouched in that range. 

This  asymmetry of the edge states when coupling with the continuum  has important consequences in the transport properties of the system as we will discuss in the following sections, for it affects asymmetrically edge states traveling in opposite directions, which determine both directions of transport in a two-terminal set-up.

%%%%%%%%%%%%%%%%%%%%%%%%%%%%%%%%%%%%%%%%%%%%%%%%%%%%%%%%%
\section{Two Terminal Conductance\label{twoterminal}}%%%%
%%%%%%%%%%%%%%%%%%%%%%%%%%%%%%%%%%%%%%%%%%%%%%%%%%%%%%%%%

%%%%%%%%%%%%%%%%%%%
\begin{figure}[b]
\includegraphics[width=.9\columnwidth]{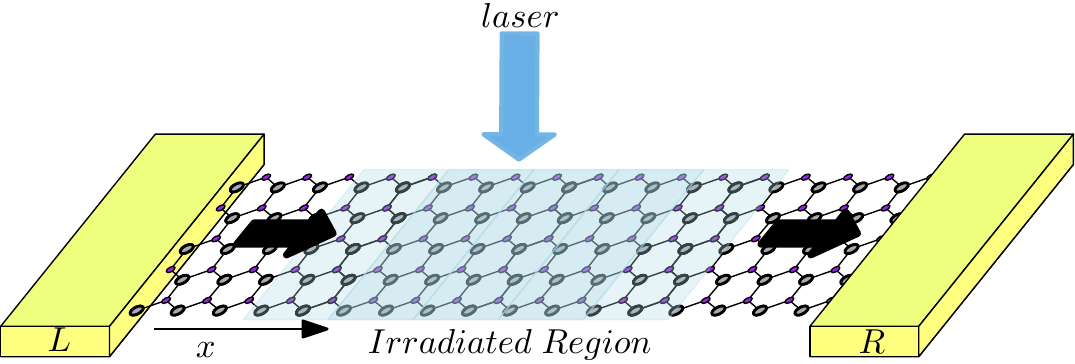}
\caption{(Color online) Setup for the calculation of the conductance using the Landauer-Buttiker approach for Floquet systems \cite{Arrachea2006,FoaTorres2014}. The system (including the leads) is WS$_2$ and the laser field is applied along the central region. The intensity of the field is decreased smoothly towards the leads until it vanishes\label{set-up}.}
\end{figure}
%%%%%%%%%%%%%%%%%%%%

In this section we present results for the two-terminal conductance of a WS$_2$ ribbon in the presence of a time periodic driving. For that purpose we separate the ribbon into three regions (Fig.~\ref{set-up}): a central region where the laser field is turned on (see details below) and two regions on the left and the right that have no driving and constitute the source and drain reservoirs (leads), respectively. The calculation is done using the Green function technique for the calculation of the transmission coefficient within the Landauer-Buttiker formalism \cite{Baranger1989,Pastawski1992,Pastawski2001} properly adapted to the present case using Floquet theory~\cite{Moskalets2002,Camalet2003a,Arrachea2006,Kohler2005}. 

To start with we calculate the transmission (the linear conductance is proportional to it) in the absence of the laser field. The result is shown in Fig.~\ref{equiT} for the particular case of a zigzag ribbon.  As expected, it reproduces the features of the band structure shown in Fig.~\ref{zigzag}. Notice in particular that the presence of the equilibrium edge states inside the bulk gap leads to a constant transmission of $2$ in that energy range due to the presence of two distinct edge states (one on the W-edge and one on the S-edge)---the factor $2$ coming from the spin variable is not included in the transmission but added at the end to the expression for the current (see Eq.~\eqref{current1}).

%%%%%%%%%%%%%%%%%%%%%%%%%%%%%%%%%%%%%%%%%%%%%%%%%%%%%%%%%%%%%%%%%%%%
\subsection{Effects of the Floquet Gaps and Floquet Edge States}%%%%
%%%%%%%%%%%%%%%%%%%%%%%%%%%%%%%%%%%%%%%%%%%%%%%%%%%%%%%%%%%%%%%%%%%%
We now consider the time dependent case. The set-up is depicted in Fig.~\ref{set-up}.  The amplitude of the laser field ($A_0$) is taken to be constant inside the central region and it slowly switches off near the leads, where it becomes zero. This is modeled by defining a local parameter $s$  near the leads as $s(x)=\frac{eaA_0}{4\hbar c} (1\pm\cos( \pi x/\lambda))$, where $x$ is the spatial coordinate along the ribbon, the minus/plus sign corresponds to the region near the left/right lead and $\lambda$ defines the length (along the ribbon) of the switching region. Once $s$ reaches its maximum value ($eaA_0/2\hbar c$) it is kept constant between leads.
Since the numerical calculation requires the use of large matrices---the effective width of the ribbon is augmented by the number of Floquet replicas as well as the three orbitals involved---they become quite demanding for large systems. For that reason we work with ribbons up to $130$ tungsten atoms wide, which are large enough to shown the main features that the laser field introduces into the transport properties. 
%%%%%%%%%%%%%%%%%%%%%%%%%%%%%%%%%
\begin{figure}[t]
\includegraphics[width=0.95\columnwidth]{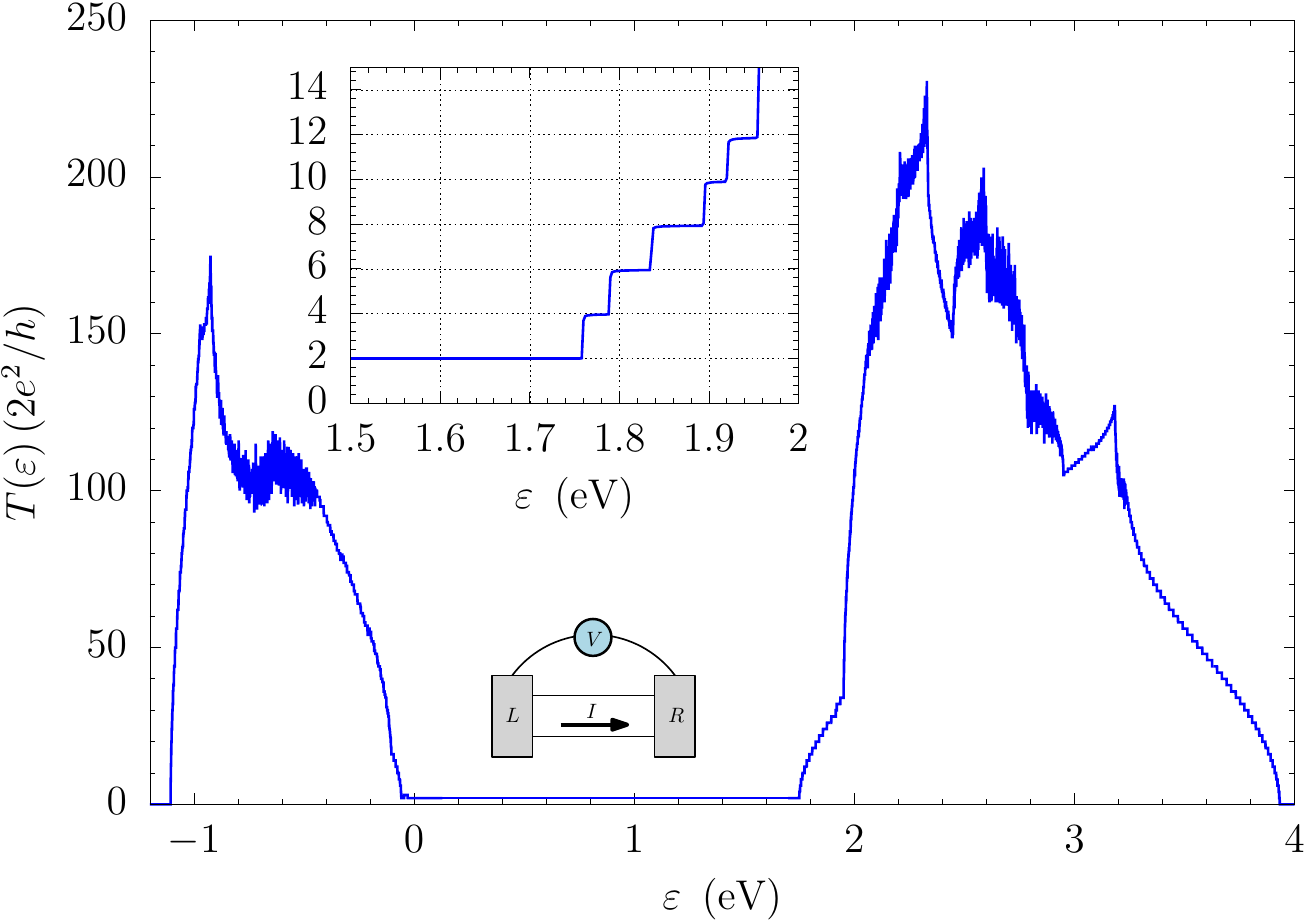}
\caption{(Color online) Non irradiated two-terminal transmittance of a zigzag  WS$_2$ nano ribbon (130 tungsten atoms wide) at zero temperature. The non zero values along the semiconductor gap (roughly ${0\leqslant \varepsilon\leqslant 2\,\mbox{eV}}$) are due to the two equilibrium edge states traveling in each direction (see Fig.~\ref{zigzag}).  The inset shows the transmittance in the region ${1.5\,\mbox{eV}\leqslant \varepsilon\leqslant 2\,\mbox{eV}}$, where it is clearly monotonous \label{equiT}.}
\end{figure}
%%%%%%%%%%%%%%%%%%%%%%%%%%%%%%%%%

The time-averaged two-terminal conductance can be written as \cite{FoaTorres2014}
%%%%%%%%%%%%%%%%%%%%%%
\begin{eqnarray}\label{current1}
\nonumber
\bar{I}&=&\frac{1}{T}\int_0^T dt\, I(t)\,,\\
\bar{I}&=&\frac{2e}{h}\sum_n \int d\varepsilon\, \left[T_{RL}^{(n)}\,f_L(\varepsilon) - T_{LR}^{(n)}\,f_R(\varepsilon)\right],
\end{eqnarray}
%%%%%%%%%%%%%%%%%%%%%%%
where $T_{RL}^{(n)}(\varepsilon)$ is the transmission probability for an electron from lead $L$ with energy $\varepsilon$  to lead $R$ emitting (absorbing) $n>0$ ($n<0$) photons and $f_{\alpha}(\varepsilon)$ is the Fermi functions at lead $\alpha$ ($\alpha=L,R$). In the absence of many-body interactions this is equivalent to the Keldysh formalism \cite{Kohler2005,Arrachea2006}. 
 Defining the quantities $T(\varepsilon)=1/2\sum_n\,\left(T_{LR}(\varepsilon)+T_{RL}(\varepsilon)\right)$ and $\delta T(\varepsilon)=1/2\sum_n(T_{LR}(\varepsilon)-T_{RL}(\varepsilon))$, the current $\bar{I}$ can be written as the sum of two terms
%%%%%%%%%%%%%%%%%%%%
\begin{equation}
\bar{I}\!=\!\frac{2e}{h}\!\int\! d\varepsilon \big[ T(\varepsilon)(f_L(\varepsilon)\!-\!f_R(\varepsilon))
     \!+\!\delta T(\varepsilon)(f_L(\varepsilon)\!+\!f_R(\varepsilon))\big]\,.
\end{equation}
%%%%%%%%%%%%%%%%%%%%
\begin{figure}[t]
\includegraphics[width=0.95\columnwidth]{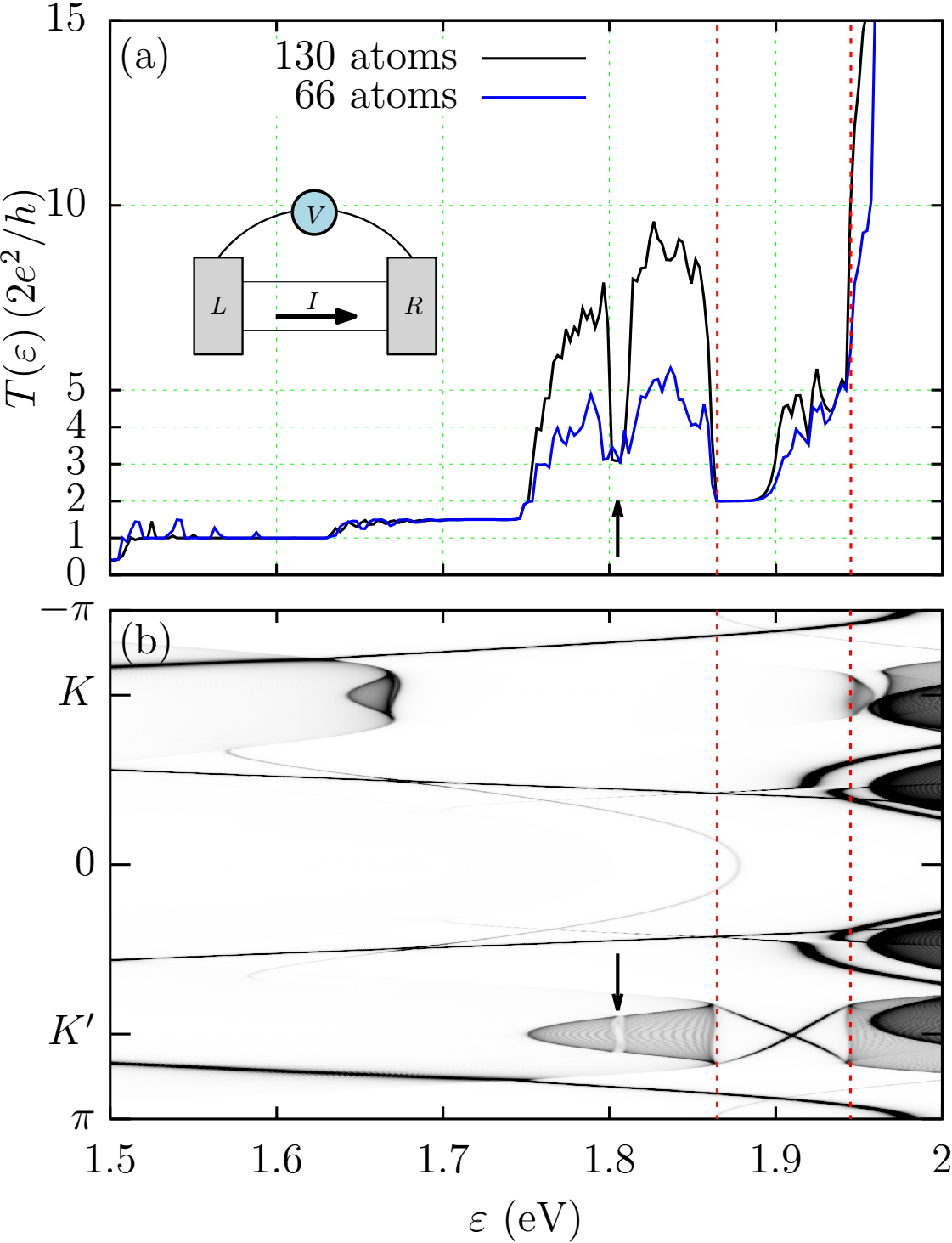}
\caption{(a) Transmittance of an irradiated zigzag ribbon in the red detuned regime ($\hbar\Omega=1.91\,$eV and $s=0.04$) for two different widths (60 and 130 W atoms wide). In both cases we see clear signatures of suppression in the energy ranges of the Floquet gaps. Interestingly, there are energy regions where the conductance does not depend on the ribbon's width, and so they must correspond to electron transport due to edge states. The vertical dashed red lines indicate the dynamical gap. (b) $0$-Floquet bands, for the same parameters $\hbar\Omega$ and $s$ as in (a) but for a large width (same as Fig.~\ref{zoomFloquet}). The black arrow in both figures correspond to the small gap that appears in the crossing of $|c,0\rangle$ and $|v,1\rangle$ replicas \label{figure_red}.}
\end{figure}
%%%%%%%%%%%%%%%%%%%%%%%%%%%%
Keeping only linear terms in the bias voltage $\delta V$ and con\-si\-de\-ring  the zero temperature limit (this is not a limitation and can be easily generalized) the above expression reduces to
%%%%%%%%%%%%%%%%%%%
\begin{equation}
\bar{I}=\frac{2e^2}{h}\,T(\varepsilon_F)\,\delta V+\frac{4e}{h}\int_{-\infty}^{\varepsilon_F} d\varepsilon\, \delta T(\varepsilon) \,.
\end{equation}
%%%%%%%%%%%%%%%%%%
The last term in this equation is the so-called {\it pumped current} and it appears when there is an asymmetry in the transmission probability, that is $T_{LR} \neq T_{RL}$, and  it is clear that it does not depend on the applied bias and so it is present even in the absence of a voltage drop. The mirror symmetry in an armchair ribbon (see Sec.~\ref{EqEdgeStates}) leads  to $T_{LR} = T_{RL}$, and thus in this kind of ribbons there is no pumped current.
%%%%%%%%%%%%%%%%%%
\begin{figure}[t]
\includegraphics[width=0.95\columnwidth]{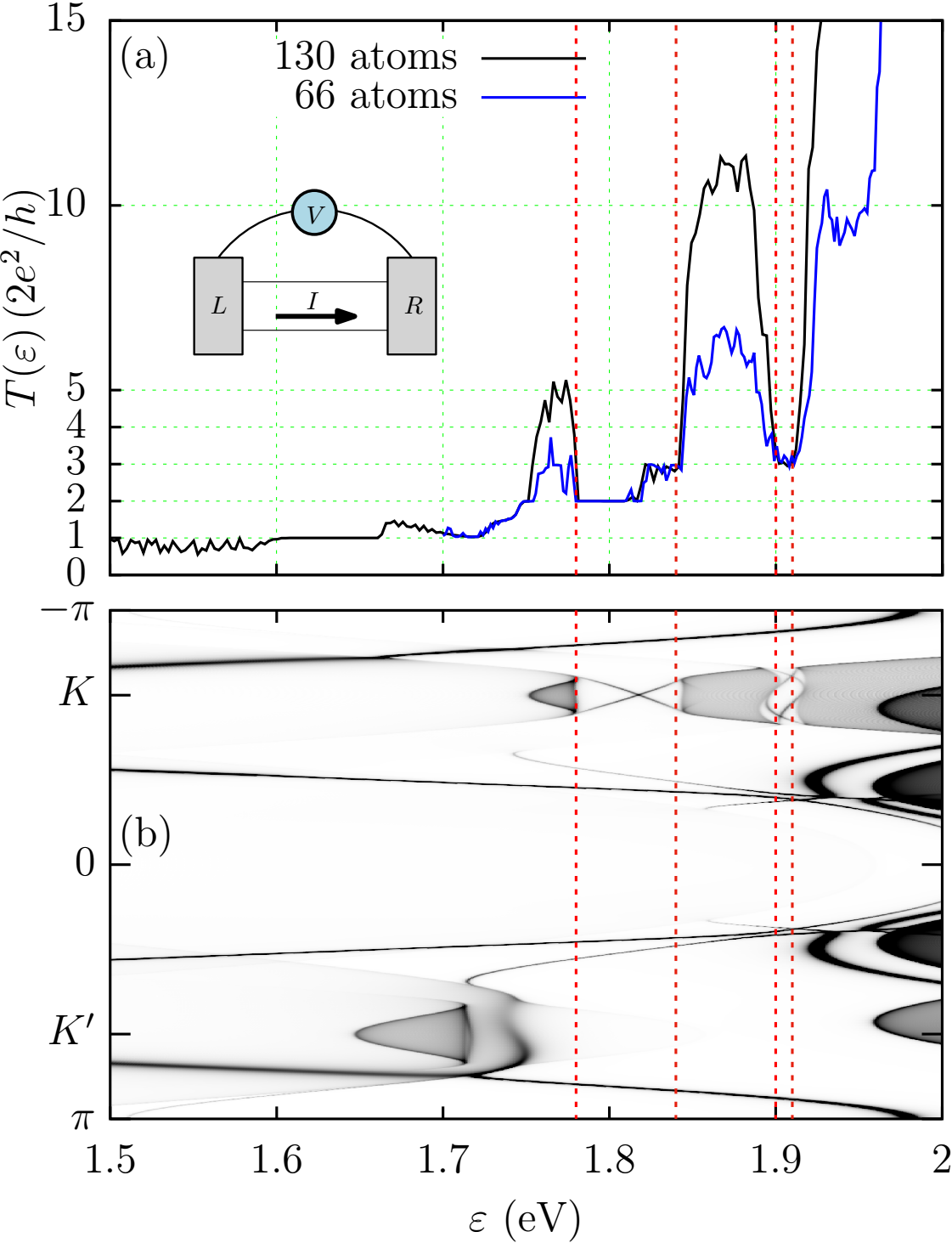}
\caption{Same as in the previous figure but for the  blue detuned regime ($\hbar\Omega=2.07\,$eV and $s=0.05$). Both dynamical gaps are marked with dashed red lines. Notice that the features in the lower gap (fomed by the couple of $|c,0\rangle$ and $|x,-1\rangle$) are similar to those in Fig.~\ref{figure_red}
\label{figure_blue}.}
\end{figure}
%%%%%%%%%%%%%%%%%%%%%%%%%%%%%

Figs.~\ref{figure_red} and~\ref{figure_blue} show the transmittance $T(\varepsilon)$  for two zigzag ribbons of different widths ($66$ and $130$ tungsten atoms) in the red ($\hbar\Omega=1.91\,$eV and $s=0.04$) and blue ($\hbar\Omega=2.07\,$eV and $s=0.05$) detuned regimes, respectively.  In order to  describe correctly the Floquet gaps we allow the switching on and off of the laser intensity to take place over a length of $100$ tungsten atoms, which corresponds to $\lambda=100a$ ($a$ being the lattice constant), the homogeneous central region being also of $100$ atoms in length (along the ribbon). This reduces the effect  of multiple reflections at the entrance and exit of the irradiated region (which appear as  Fabry-Perot type oscillations) on the conductance. The $0$-Floquet bands are also shown in the corresponding energy range for the purposes of comparison and identification of the relevant features : Floquet gaps and their corresponding Floquet edge states. Interestingly, there are quasienergy regions where the transmittance clearly does not depend on the ribbon's width, which is a hint that for this values of quasienergy the electron transport occurs entirely through edge states. On the other hand, as expected, there are dips in $T(\varepsilon)$ coinciding with the quasienergy ranges of the Floquet gaps, as compared with the monotonous behaviour in the non irradiated case (see inset in Fig.~\ref{equiT}). There, the transmittance is significantly reduced but it does not drop completely to zero due to the presence of Floquet edge states---this is consistent with the fact that the transmittance is almost the same for the two widths--the lack of exact quantization of the conductance in the Floquet case is to be expected, as it has been discussed previously~\cite{FoaTorres2014,Farrell2015}. Yet, in this case the analysis of the number of edge states involved becomes somehow intrincate due to the fact that, apart from the Floquet edge states inside the dynamical gaps, there are equilibrium edge states already present in the absence of radiation---this represents a departure from graphene case where only Floquet edge states matter~\cite{FoaTorres2014}. 

It is worth emphasizing that the difference in size of the two Floquet gaps shown in Fig.~\ref{14} together with the reduction of the transmittance induced by them, allows in principle to identify the topological transition with a transport measurement.  To this end, we note that the red detuned gap is in general wider than the blue detuned one. For quasienergies below the center of the red detuned dynamical gap, the transmittance is identical to that of the non irradiated sample ($T(\varepsilon)=2$), indicating a poor matching between the incoming wave function and the Floquet edge states inside the irradiated region~\cite{FoaTorres2014,Farrell2015}. Above the center of the gap we can see the influence of the chiral edge states, although the quantization is not recovered due also to the matching problem and the presence of equilibrium edge states. The blue detuned gap is significantly smaller  and the transmittance there has a mean value roughly equal to $T(\varepsilon)=3$ in both regimes. 

%%%%%%%%%%%%%%%%%%%%%%%%%%%%%%%%%%%%%%%%%%%%%%%%%%
\begin{figure}[tpb]
\includegraphics[width=0.8\columnwidth]{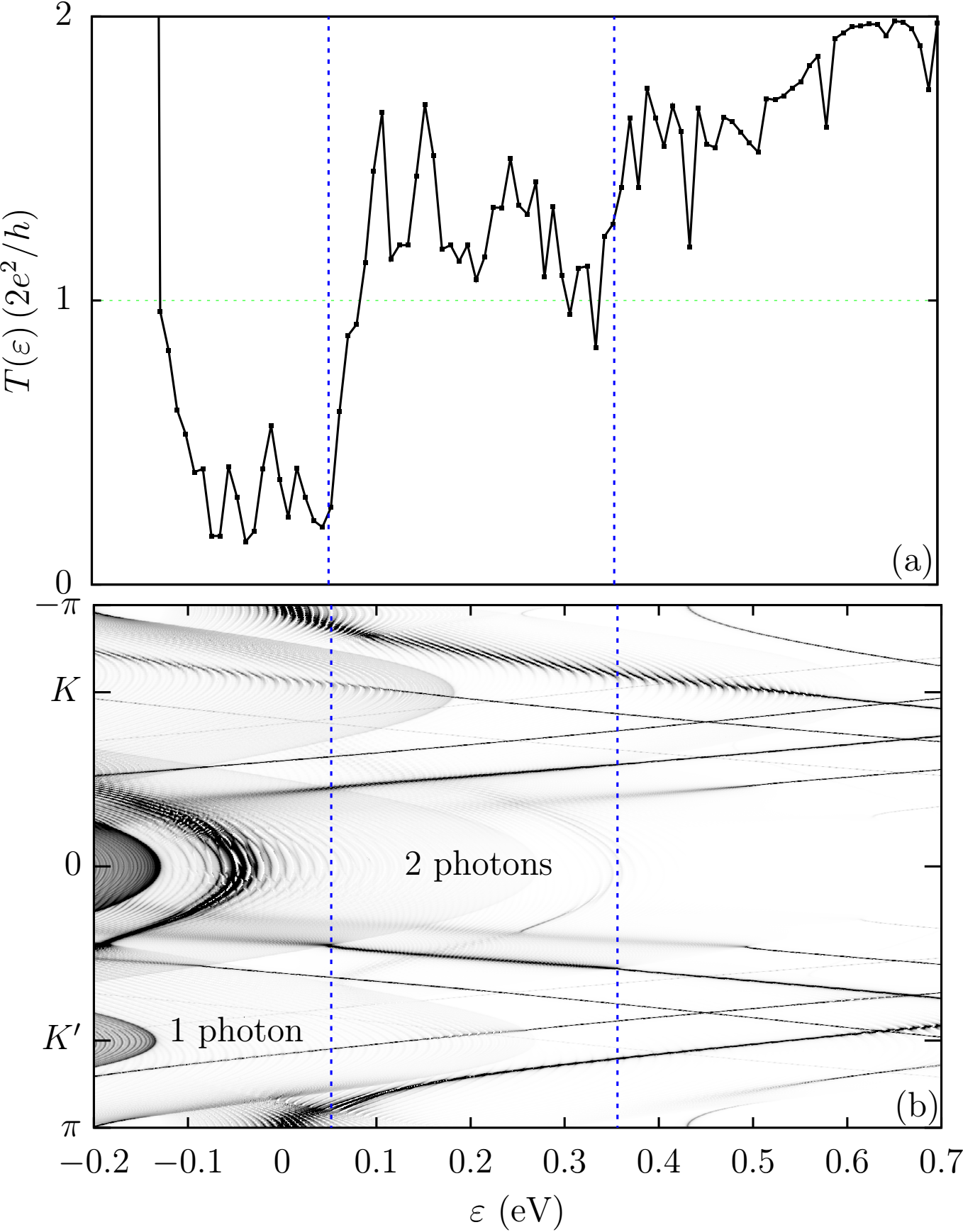}
\caption{(a) Transmittance $T(\varepsilon)$ for $\hbar\Omega=0.4\,$eV and $s=0.18$ of a 130 tungsten atoms wide ribbon. The vertical blue dashed lines separate quasienergy regions where the photonics processes involve one or two photons. (b)  $0$-Floquet bands.  The transmittance is the smallest where the equilibrium edge states couples with the $m=1$ replicas of the valence band, whereas where the coupling is with the $m=2$ the suppression in transmittance is less important \label{mess}.}
\end{figure}
%%%%%%%%%%%%%%%%%%%%%%%%%%%%%%%%%%%%%%%%%%%%%%%%%%%%%%%%%%%%%%

Another important aspect is that there is a suppression of the conductance due to the equilibrium edge states. The magnitude of such suppression is in  close relationship with the number of photons exchanged between the different replicas. This effect is more clearly seen if  we change to the regime given by $\hbar\Omega=0.4\,$eV and $s=0.18$. As we know, with zero laser intensity ($s=0$), the conductance along the semiconductor gap in the zigzag ribbon is due entirely to the  equilibrium edge states  that are shown in Fig.~\ref{zigzag}. In units of $2e^2/h$ this accounts for a constant transmittance of two ($T(\varepsilon)=2$).   In Fig.~\ref{mess}(a) the transmittance $T(\varepsilon)$ of a $130$ tungsten atoms wide ribbon is shown, along with the $0$-Floquet bands of the same ribbon in the same quasienergy range (Fig.~\ref{mess}(b)). In Fig.~\ref{mess}(b) the region where the equilibrium edge states couple with the $m=1$ and $m=2$ replicas of the valence bands are shown, and the same regions are marked in Fig.~\ref{mess}(a). This clearly shows that the process involving only one photon  is more effective in suppressing the conductance, and that this effect is less important as the order $|m|$ of the Floquet processes increases. The noise-like behaviour that Fig.~\ref{mess} exhibits results from the coupling between transverse modes of the Floquet bulk replicas with the equilibrium edge states. These transverse modes have their origin in the smallness of the ribbon's width and can be seen in Fig.~\ref{mess}(b) as anticrossings in the $0$-Floquet bands.

%%%%%%%%%%%%%%%%%%%%%%%%%%%%%%%%%%%%%%%%%%%%%%%%%%%%%%%%%%%%%%
\subsection{Left/Right Asymmetry: Pumped Current}%%%%%%%%%%%%%
\begin{figure}[t]
\includegraphics[width=0.9\columnwidth]{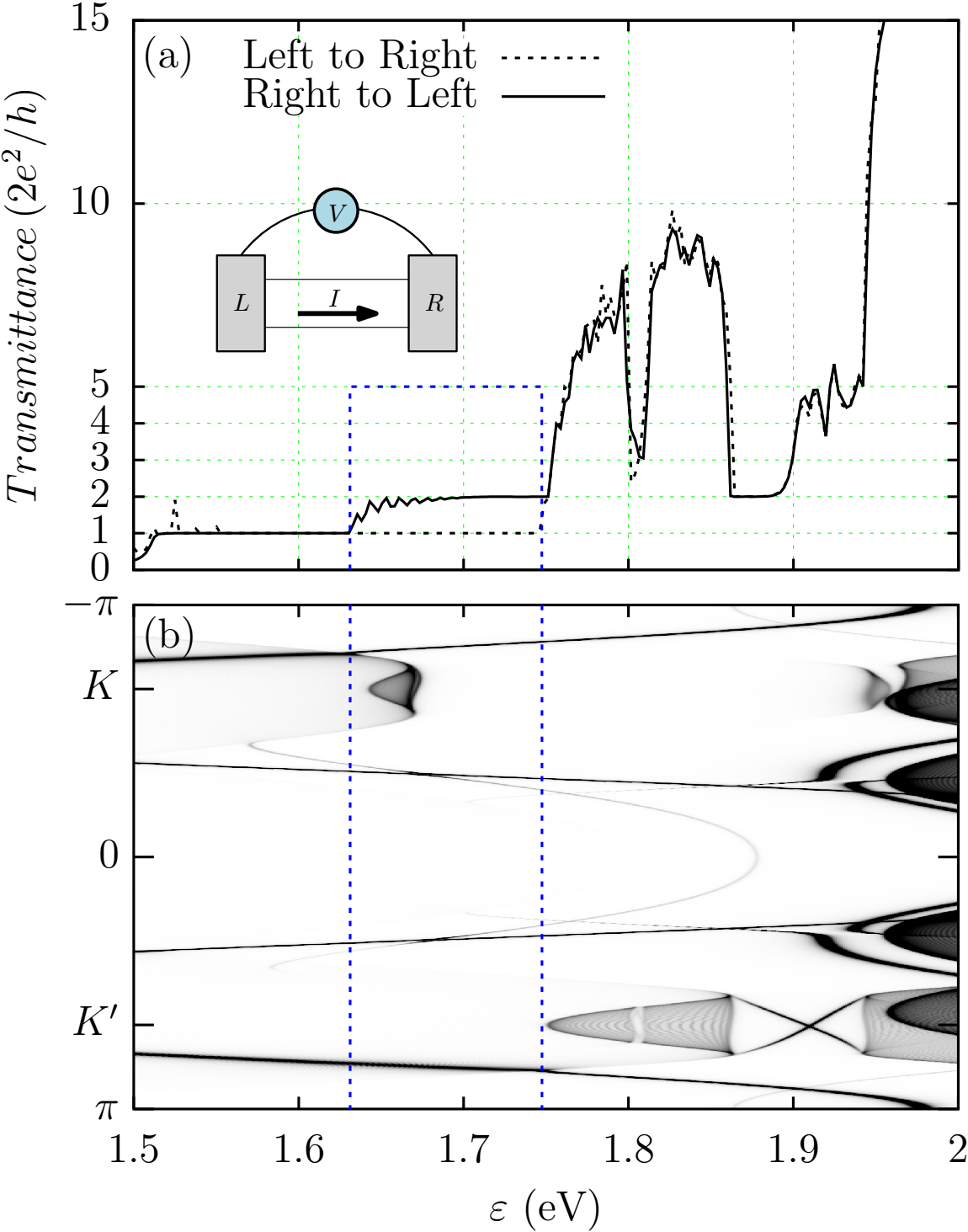}
\caption{(Color online) (a) Zero temperature transmittance of a zigzag ribbon for both directions of electron flow in the red detuned case ($\hbar\Omega=1.91\,$eV, $s=0.04$). For energies below roughly  1.75$\,$eV, the transmittance  is through edge states. In the energy region between the vertical dashed blue lines, the conductance goes from 2 to 1 when changing the flux direction, remaining roughly the same in the rest of the energy interval shown. (b) $0$-Floquet density of states for the same ribbon (bulk and both edges). The S-edge states  contribute with a transmittance 1 in every direction. In the region between the horizontal dashed blue lines, the W-R edge state  is supressed due to the coupling with the continuum of a replica, while the W-L is not. This produces the asymmetry in conductance seen in (a) \label{figure_directional}.}
\end{figure}

%%%%%%%%%%%%%%%%%%%%%%%%%%%%%%%%%%%%%%%%%%%%%%%%%%%%%%%%%%%%%%
We now discuss the lack of left/right symmetry of the transmittance, ${T_{LR}\neq T_{RL}}$, as shown in Fig.~\ref{figure_directional}. While the asymmetry is present on the entire energy range (with different magnitudes), we concentrate here on the quasienergy range corresponding to the equilibrium semiconductor gap, for there the effect is more easily seen. This same phenomenon has already been addressed in Sec.~\ref{effectonequilibrium} for the $0$-Floquet spectral density.  Fig.~\ref{figure_directional}(a) shows the transmittance for each direction ($L\rightarrow R$ and $R\rightarrow L$)) for a zigzag ribbon in the red detuned case (parameters $\hbar\Omega=1.91\,$eV and $s=0.04$) near the bottom of the conduction band. It is clear from this figure that there is a large right/left asymmetry in the region delimited by the vertical dashed blue lines. Namely, there is one transport channel difference between the two directions. This feature can be easily understood by looking at the $0$-Floquet bands shown in Fig.~\ref{figure_directional}(b) (this is the same as Fig.\ref{zoomFloquet} but rotated). First, we point out again that there are four edge states, that we named as the W- and S- edge states, that run in both directions (positive and negative slop as a function of $k_x$, see Figs.~\ref{zigzag} and \ref{zoomFloquet} for a better reference). As we discussed previously, the S-edge states  are almost unaffected by the radiation in that quasienergy range since they couple weakly to it, so they contribute with a factor $\sim1$ equally to both directions, as in equilibrium. On the contrary, the W-edge states couple strongly to the bulk Floquet replicas, in that quasienergy region, and hence their contribution to the transmission is altered with respect to the equilibrium case. In particular, due to the differences on the optical Stark shift in each one of the $K$-points, the splitting of the bulk bands is quite different in $K$ and $K'$ points and so is the quasienergy region where they overlap (and couple) with the W- edge state. This manifests particularly  as a broadening  of the W-state traveling to the right (state W-R in Fig.~\ref{zoomFloquet}), while the W-state traveling to the left (state W-L) is almost unaffected. This broadened right-moving W-state does not contribute to the transmittance (or it does with a significantly smaller value), whereas the left-moving states does it with . This results in the strong asymmetry $\sum_nT^{(n)}_{LR} \neq \sum_nT^{(n)}_{RL}$ shown in Fig.~\ref{figure_directional}. The sharp difference $\sum_nT^{(n)}_{LR} - \sum_nT^{(n)}_{RL}\sim1$ found here comes as a result of the particular value of $s$ chosen and it might be smaller for other values. It is worth mentioning that similar asymmetries has been found in other systems such as a topological insulator coupled with a metal~\cite{FoaTorres2016} and other Floquet systems such as irradiated bilayer graphene~\cite{DalLago2017}. Because the suppression occurs in the W-edges, it is clear that the current flow in that regime is not homogeneous, being larger on the S-edge (whose edge states are not affected).

%%%%%%%%%%%%%%%%%%%%%%%%%%%%
%%%%%%%%%%%%%%%%%%%%%%%%%%%%

\begin{figure}[t]
\includegraphics[width=0.9\columnwidth]{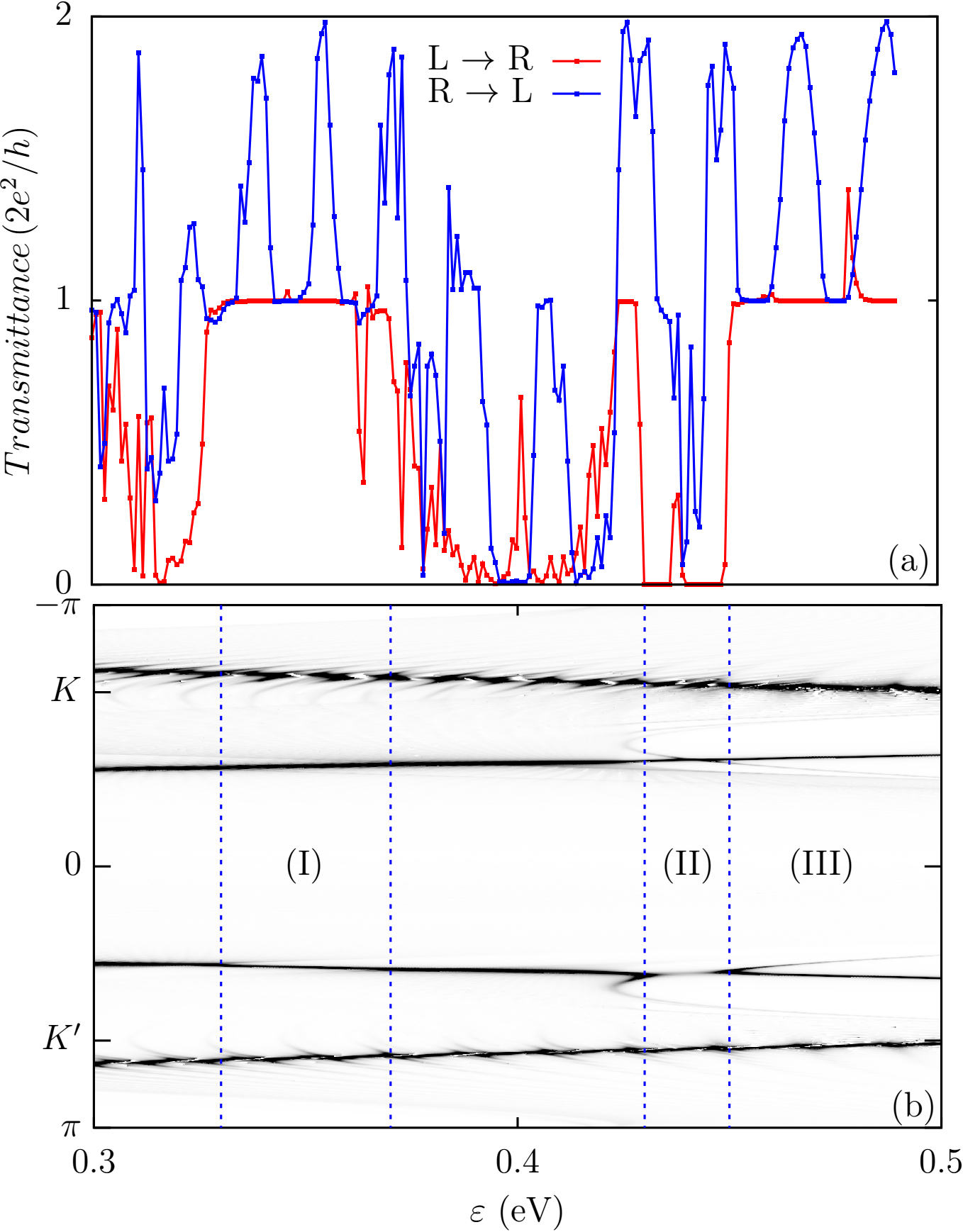}
\caption{(Color online) (a) Transmittance in the red detuned regime ($\hbar\Omega=1.91\,$eV, $s=0.04$) in the quasienergy interval ${0.3\, \mbox{eV}\leqslant\varepsilon\leqslant 0.5\,\mbox{eV}}$ of a $130$ atoms wide ribbon for both directions.  (b) $0$-Floquet bands. The region (I) only one right-traveling is not wixed with the continuum and thus contributes to the transmittance. In (II) a gap appears as a result to the coupling with another edge state and the transmittance drops to zero.  Note that exchange of roles of the W- and S-edge states with respect to the previous case (the W-edge states are now closer to $k=0$).\label{mess_gaps}}
\end{figure}

A similar asymmetry can be found when examining the quasienergy region ${0.3\,\mbox{eV}\leqslant \varepsilon \leqslant 0.5\,\mbox{eV}}$ (the metallic zigag equilibrium edge states). This is shown in Fig~\ref{mess_gaps}, where the top panel corresponds to the transmission coefficient  $T(\varepsilon)$ while the bottom one is the  $0$-Floquet bands in the corresponding quasienergy range. The width of the ribbon in both calculations is the same (130 tungsten atoms). The regime is red detuned with the same parameters as Fig.~\ref{figure_red}. Here, in addition to the broadening induced by the coupling with the bulk states, there is a dynamical gap (at roughly $\varepsilon=0.
43\,$eV) in the quasienergy dispersion of the W-edge states originated by the mixing with the replica of the other W-edge near the bottom of the conduction band (cf. Fig.~\ref{irradiated-edges}).  This completely degrades the transmission in one direction as seen in Fig.~\ref{mess_gaps}(a). This is indicated by region (II) between vertical dashed blue lines. The suppression in conductance around $\varepsilon=0.4\,$eV occurs due to coupling with bulk states. This clearly shows that the Floquet replicas are highly effective  in suppressing the static edge conductance.

Notice also that there are regions of quasienergy where the edge states (in a given direction) are not coupled to the Floquet replicas and hence the transmittance reaches its maximum value of one, indicated by regions (I) and (III). On the other hand, and as before, the noisy behaviour is a finite size effect related to the quantization of the bulk states (transverse modes) along a direction transversal to the ribbon's length. This could be eliminated by using a wider ribbon, but this increases dramatically the computational time and it is beyond our capabilities. In any case, we expect the suppression of the transmittance to be more homogeneous for larger ribbons.

We emphasize again that this asymmetry is absent in the case of armchair ribbons due to their reflection symmetry (see Sec.~\ref{EqEdgeStates}).  

%%%%%%%%%%%%%%%%%%%%%%%%%%%%%%%%%%%%%%%%%%%
%%%%%%%%%%%%%%%%%%%%%%%%%%%%%%%%%%%%%%%%%%%%%%%%%%

%%%%%%%%%%%%%%%%%%%%%%%%%%%%%%%%%%%%%%%%%%%%%
\section{Conclusions \label{conclusions}}%%%
%%%%%%%%%%%%%%%%%%%%%%%%%%%%%%%%%%%%%%%%%%%%%
In this paper, we have studied how the electronic structure of a monolayer TMDC ribbon (taking WS$_2$ as an example) is affected by a monocromatic laser field using the framework given by Floquet theory and a three-band tight binding model developed by Liu {\it et al}~\cite{liu2013}, with hopping terms up to third nearest neighbours.

Our results of the the bulk Floquet bands are consistent with previous works \cite{Claassen2016}. The optical selection rules give place to an asymmetrical gap opening in both high symmetry $K$-points, which can be explained (and quantified to lowest order at least) by looking at the Floquet matrix around each of these points and reducing the whole extended Hilbert space $\mathcal{T}\otimes \mathcal{R}$ to the appropriate number of Floquet replicas (see Appendix C). This asymmetry has a deep significance in understanding all the features we found in the $0$-Floquet spectrum and in the two-terminal conductance.
Apart from the expected emergence of chiral Floquet edge states (as is the case with graphene), and the topological transition that appears when going from the red to the blue detuned regime, we find that, in the case of zigzag ribbons, there are also very unusual and interesting effects on the equilibrium edge states. Namely, we find that small gaps can develop at points where there is a resonance between them and those edge states near the bottom of the $c$- and $x$-bands or the top of the $v$-band. When the coupling involves bulk states, it leads to a broadening of the formerly peaked spectral density of the edge state (and hence a loss of  spectral weight in the $0$-Floquet bands). The magnitude of this broadening depends on the number of photons involved in the process (the quantum number $m$ of the replica), and it can be identified as a different level of suppression in the two-terminal transmittance.
These effects are different for the different edge states and differ for those traveling to the right or to the left depending on the polarization (clockwise or anticlockwise) of the laser field---which is responsible for the breaking of time reversal symmetry.

In addition, we performed two-terminal transport calculations in the presence of a laser field, and compare our results with the Floquet bands previously found. On the one hand,  we verify that the Floquet gaps found earlier show up as dips in the conductance in the correct quasienergy interval. The value of the conductance inside these gaps is not zero due to the contribution of the chiral edge states and although its value is not quantized it  is seen to be  independent of the ribbon's width, which is a signature of  edge transport. On the other hand, we found that
the asymmetry between the $K$ and $K'$ points in the Floquet bands cause the transmittance to exhibit  different values for each direction of the electron flux for the zigzag case, leading to a pumped current without bias. This effect is particularly important when transport involves the equilibrium edge states found in zigzag ribbons, and can even lead to a switch off of the conductance (in a given direction) depending on the laser amplitude. This offers an interesting prospect for future research, specially on the light of the recent experimental results on light-induced anomalous Hall effect in graphene \cite{McIver2018}.
%%%%%%%%%%%%%%%%%%%%%%%%%%%%%%%%%%% 
\begin{acknowledgments}
We acknowledge financial support from ANPCyT (grants PICTs 2013-1045 and 2016-0791), from CONICET (grant PIP 11220150100506) and from SeCyT-UNCuyo (grant 06/C526).
\end{acknowledgments}
%%%%%%%%%%%%%%%%%%%%%%%%%%%%%%%

\appendix
\section{The Three-Band Model for TMDC \label{band-model}}
In this section we present, for the sake of completeness, the three-band model for TMDC developped by Liu {\it et al}~\cite{liu2013}, with hoppings up to third nearest neighbours (a model with only nearest neighbours has also been presented although it was shown to describe correctly the bands only in a small region around $K$ and $K'$ points). This model uses the atomic bases $\{d_{z^2},d_{xy},d_{x^2}\}$ of the tungsten atom alone, something that is known to be sufficient to correctly describe the valence and conduction bands. The Hamiltonian will be constructed on the basis of the  following Bloch wavefunctions:
\begin{align}\label{blochwaves}
|d_{z^2}(\bm{ k})\rangle &= \frac{1}{\sqrt{N}}\,\sum_{\bm{R}}\,e^{i\bm{ k}\cdot\bm{R}}\,|d_{z^2}(\bm{R})\rangle,  \notag\\
|d_{xy}(\bm{ k})\rangle  &= \frac{1}{\sqrt{N}}\,\sum_{\bm{R}}\,e^{i\bm{ k}\cdot\bm{R}}\,|d_{xy}(\bm{R})\rangle ,  \notag\\
|d_{x^2}(\bm{ k})\rangle &= \frac{1}{\sqrt{N}}\,\sum_{\bm{R}}\,e^{i\bm{ k}\cdot\bm{R}}\,|d_{x^2}(\bm{R})\rangle.
\end{align}

The symmetries of the Hamiltonian can be used to show that the numbers of independent hoppings between orbitals at different sites are six for the nearest neighbours, five for the second neighbours and six for the third neighbours. These hoppings are denoted $t_i$, $r_i$ and $u_i$ respectively, and their values are obtained by using DFT~\cite{liu2013}. The final Hamiltonian $H(\bm{ k})$ can be written in the following form:
\begin{equation}\label{hamilton}
H(\bm{ k})=\left(
\begin{array}{ccc}
 h_{11} & h_{12} & h_{13} \\
 h_{12}^* & h_{22} & h_{23} \\
 h_{13}^* & h_{23}^* & h_{33} 
\end{array} \right).
\end{equation}
where the matrix elements are given by
\begin{widetext}
\begin{eqnarray}
h_{11} &= & \varepsilon_1 + 2t_0(2\cos\alpha\cos\beta+\cos2\beta)
 + 2r_0(2\cos3\alpha\cos\beta+\cos2\beta)               
 + 2u_0(2\cos2\alpha\cos2\beta+\cos4\alpha)\,,                \\
h_{22} &= & \varepsilon_2 + (t_{11}+3t_{22})\cos\alpha\cos\beta + 2t_{11}\cos2\alpha 
 +4r_{11}\cos3\alpha\cos\beta + 2(r_{11}+\sqrt{3}r_{12})\cos2\beta \notag\\
 &&+ (u_{11}+3u_{22})\cos2\alpha\cos2\beta + 2u_{11}\cos4\alpha\,, \\
h_{33} &= & \varepsilon_2 + (3t_{11}+t_{22})\cos\alpha\cos\beta + 2t_{22}\cos2\alpha     
       + 2r_{11}(2\cos3\alpha\cos\beta+\cos2\beta)                             \notag \\      
       &&+\frac{2}{\sqrt{3}}r_{12}(4\cos3\alpha\cos\beta-\cos2\beta)             
       + (3u_{11}+u_{22}) \cos2\alpha\cos2\beta + 2u_{22}\cos4\alpha\,,\\            
h_{12} &=& -2\sqrt{3}t_2\sin\alpha\sin\beta + 2(r_1+r_2)\sin3\alpha\sin\beta             
          -2\sqrt{3}u_2\sin2\alpha\sin2\beta                                           \notag\\
          &&+2it_1\sin\alpha(2\cos\alpha+\cos\beta)+2i(r_1-r_2)\sin3\alpha\cos\beta      
          +2iu_1\sin2\alpha(2\cos2\alpha+\cos2\beta)                                  \,, %\\
          \end{eqnarray}
\begin{eqnarray} 
h_{13} &=& 2t_2(\cos2\alpha-\cos\alpha\cos\beta)                                         
        -\frac{2}{\sqrt{3}} (r_1+r_2)(\cos3\alpha\cos\beta-\cos2\beta)
        +2i\sqrt{3}u_1\cos2\alpha\sin2\beta                   \notag\\
       && +2u_2(\cos4\alpha-\cos2\alpha\cos2\beta) +2i\sqrt{3}t_1\cos\alpha\sin\beta     
       +\frac{2i}{\sqrt{3}}(r_1-r_2)\sin\beta(\cos3\alpha+2\cos\beta)                  \,, \\
h_{23} &=& \sqrt{3}(t_{22}-t_{11}) \sin\alpha\sin\beta + 4r_{12} \sin3\alpha\sin\beta    
       +\sqrt{3}(u_{22}-u_{11})\sin2\alpha\sin2\beta                                  \notag\\
       && +4it_{12} \sin\alpha(\cos\alpha-\cos\beta)                                     
        +4iu_{12}\sin2\alpha(\cos2\alpha-\cos2\beta)  .                                 
\end{eqnarray}
\end{widetext}
Here, we defined the dimensionless parameters $\alpha=ak_x/2$ and $\beta=\sqrt{3}ak_y/2$. As it was mentioned before,
the hoppings parameters $t_i$, $r_i$ and $u_i$ (as well as the on site energies $\varepsilon_1$ and $\varepsilon_2$) have been obtained by Liu {\it et al} \cite{liu2013} by fitting the analytical eigenvalues obtained from $H({\bm k})$ with DFT bands  at the $K$ point. For the particular case of WS$_2$ we have the following values (in units of eV)

\begin{center}
\begin{tabular}{cccccccc}
\hline\hline
$t_0$ & $t_1$ & $t_2$ & $t_{11}$ & $t_{12}$ & $t_{22}$\\ 
$-0.175$ & $-0.090$ & $0.611$ & $0.043$ & $0.181$ & $0.008$ \\
\hline
 $r_0$ & $r_1$ & $r_2$ & $r_{11}$ & $r_{12}$ & $u_0$ \\ 
 $0.075$ & $-0.282$ & $0.356$ & $2.015$ & $2.014$  & $2.056$ \\
\hline
$u_1$ & $u_2$ & $u_{11}$ & $u_{12}$ & $u_{22}$ & \\
$2.045$ & $0.659$ & $3.014$ & $0.457$ & $0.478$ & \\
\hline
$\varepsilon1$ & $\varepsilon_2$  \\
$0.717$ & $1.916$  \\
\hline
\hline
\end{tabular}
\end{center}

%%%%%%%%%%%%%%%%%%%%%%%%%%%%%%%%%%%%%%%%%%%%%%%%%%%%%%%%%%%%%%%%%%%%%%%%%%%%%%%%%%%%%
\section{Optical Selection Rules}\label{OpticalSelectionRulesandMatrixElements}%%%%%%
%%%%%%%%%%%%%%%%%%%%%%%%%%%%%%%%%%%%%%%%%%%%%%%%%%%%%%%%%%%%%%%%%%%%%%%%%%%%%%%%%%%%%

In Fig.~\ref{12} we show the Floquet bands along a line joining $K$ and $K'$ points for two different values of $\hbar\Omega$. The intensity of the coupling due to the laser field is given by the dimensionless quantity $s=eaA_0/2\hbar c$, where $a$ is the lattice parameter (distance between tungsten atoms). Looking only at $K$ and $K'$ points, we can see that the effect of circularly polarized radiation is sensitive to the $K$-point under consideration; more precisely, the direction of rotation of the vector field (clockwise or anticlockwise) determines the $K$-point affected: changing the direction of rotation  makes the $K$-points change roles. This dependency on the   rotation of ${\bm A}(t)$ can be understood in the frame of Group Theory~\cite{sie}. Looking exactly at $K$-points, the symmetry group of the system is reduced to $C_3$ (plus $\sigma_h$ symmetry which has already been taken into account). It can be shown that the Bloch waves at $K$ and $K'$ points are eigenfunctions of the operator $\hat C_3$ (rotation of $2\pi/3$ around an axis normal to the monolayer). We must keep in mind, however, that this rotation affects the $\bm{r}$ or $\bm{r-R}$ argument of the Wannier orbitals in the the Bloch waves (the so-called {\it intrinsic rotation}) as well as the lattice sites $\bm R$ in the factors $\mbox{exp}(i\bm{ k}\cdot\bm{ R})$. At both $K$ and $K'$ points the conduction band is described by:
\begin{equation}\label{b1}
|c\rangle=\frac{1}{\sqrt{N}}\sum_{\bm{ R}}e^{i\bm{k} \cdot \bm{R}}|d_{z^2}(\bm{ R})\rangle ,
\end{equation}
whereas the Bloch wave at $K$-points are given by ($\tau=1$ for $K$ and $\tau=-1$ for $K'$)
\begin{align}\label{b2}
|v_\tau\rangle  &=\frac{1}{\sqrt{N}} \sum_{\bm{ R}} e^{i\bm{ k}\cdot\bm{ R}}\frac{1}{\sqrt{2}}[\,|d_{xy}(\bm{ R})\rangle-i\tau\,|d_{x^2}(\bm{ R})\rangle\,] , \notag\\
|x_\tau\rangle  &=\frac{1}{\sqrt{N}} \sum_{\bm{ R}} e^{i\bm{ k}\cdot\bm{ R}}\frac{1}{\sqrt{2}}[\,|d_{xy}(\bm{ R})\rangle+i\tau\,|d_{x^2}(\bm{ R})\rangle\,]. 
\end{align}
As it has been pointed out by several authors \cite{sie,Cao2012,Yao2008}, the effect of the rotation upon a Bloch wave depends heavily on the center of rotation chosen, although the final selection rules from here derived cannot depend on this election. In order to make things simpler, we are going to choose the center of rotation at one tungsten site, although other alternatives have been discussed \cite{Sie2014}. In making this choice we get to the following results:
\begin{eqnarray}\label{89}
\nonumber
\hat C_3|c\rangle       =|c\rangle                                ,\, 
\hat C_3|v_\tau\rangle  = e^{\tau\frac{2\pi}{3}i}|v_\tau\rangle   ,\,   
\hat C_3|x_\tau\rangle  = e^{-\tau\frac{2\pi}{3}i}|x_\tau\rangle  \,.\\
\end{eqnarray}
These properties are relevant because of the following argument. The coupling with circularly polarized radiation between states $|\lambda_i\rangle$ and $|\lambda_f\rangle$ depends ultimately on the matrix integrals $
\langle \lambda_f| \hat P_{\scriptscriptstyle \pm} |\lambda_i \rangle$, 
where $\hat P_{\scriptscriptstyle \pm}$ is the combination of momentum operators $\hat P_x \pm  i\hat P_y$ and the sign is determined by the direction of the polarization (clockwise or anticlockwise). If $|\lambda_i \rangle$ and $|\lambda_f \rangle$ are eigenfunctions of $\hat C_3$, that is, if it holds that $\hat C_3|\lambda_\nu \rangle=e^{-i m_\nu\,2\pi/3}|\lambda_\nu \rangle$ for  $\mu=i,f$, then we can write:
\begin{align}
\langle \lambda_f |\hat P_{\pm} |\lambda_i \rangle &= \langle \lambda_f |\hat C_3^{-1}\hat C_3\hat P_{\pm}\hat C_3^{-1}\hat C_3 |\lambda_i \rangle \notag\\
& = e^{i(m_f-m_i)\frac{2\pi}{3}} \langle \lambda_f |\hat C_3\hat P_{\pm}\hat C_3^{-1}|\lambda_i \rangle,
\end{align}
and using the identity $\hat C_3\hat P_{\pm}\hat C_3^{-1}=e^{\mp\frac{2\pi}{3}i}\hat P_{\pm}$ we get to the following important result:
\begin{equation}
(1-e^{i(m_f-m_i\mp 1)\frac{2\pi}{3}}) \langle \lambda_f |\hat P_{\pm}|\lambda_i \rangle=0.
\end{equation}
The equation above tells us that a necessary condition for $\langle \varphi_f |\hat P_{\pm}|\varphi_i \rangle$ be non zero is that $m_f-m_i\mp 1=3l$, $l$ being an integer. For $\sigma_{\scriptscriptstyle +}$ radiation (anticlockwise circularly polarized) the relevant operator for a transition to a Floquet state with a extra photon ($|\lambda_i,m\rangle\rightarrow |\lambda_f,m+1\rangle$) is $\hat P_{\scriptscriptstyle{-}}$, so that the condition becomes $m_f-m_i+1=3l$. Given the values in Eq.~\eqref{89}, this condition is satisfied for $\sigma_{\scriptscriptstyle +}$ radiation only for very specific transitions, as it is shown in Table \ref{selection}. It is worth emphasizing  that these selection rules differ from those valid for atomic transition (namely $\Delta m_l=\pm 1$), and this is partially due to the ambiguity in defining the azimuthal angular momentum $m_l$: since it comes from the eigenvalue $\mbox{exp}(-i m_l2\pi/3)$, it is clear that the quantity $m_l+3k$ ($k$ integer) is also a valid angular momentum. This in turn comes from the reduced rotational symmetry of the system ($C_3$), in contrast with that of an electron in a central field (SO$(3)$).  

\begin{center}
\begin{table}
\caption{$m_f-m_i+1$ for the transition $|\lambda_i,m\rangle$ to $|\lambda_f,m+1\rangle$.}\label{selection}
\begin{tabular}{|c|ccc|ccc|}
\hline 
                & & $K$   & &  & $K'$ &  \\
       \hline
                & $|c,m\rangle$ & $|v,m\rangle$  & $|x,m\rangle$ & $|c,m\rangle$  & $|v,m\rangle$ & $|x,m\rangle$\\
       \hline
$|c,m+1\rangle$ & 1  & 2 & 0  & 1 & 0   & 2 \\
$|v,m+1\rangle$ & 0  & 1 & 1  & 2 & 1   & 3 \\
$|x,m+1\rangle$ & 2  & 3 & 1  & 0 & -1  & 1 \\
\hline
\end{tabular}
\end{table}
\end{center}

\begin{center}
\begin{table}
\caption{Bloch functions at $K$-points.}\label{functions}
\begin{tabular}{|c|c|c|}
\hline 
       &  $K$                    & $K'$                    \\
       \hline
$x$-band & $(|d_{xy}\rangle+i\,|d_{x^2}\rangle)/\sqrt{2}$  & $(|d_{xy}\rangle-i\,|d_{x^2}\rangle)/\sqrt{2}$ \\
$c$-band & $|d_{z^2}\rangle$                               & $|d_{z^2}\rangle$                              \\
$v$-band & $(|d_{xy}\rangle-i\,|d_{x^2}\rangle)/\sqrt{2}$  & $(|d_{xy}\rangle+i\,|d_{x^2}\rangle)/\sqrt{2}$ \\
\hline
\end{tabular}
\end{table}
\end{center}

%%%%%%%%%%%%%%%%%%%%%%%%%%%%%%%%%%%%%%%%
\section{Optical Stark Shift\label{StarkShift}}%
%%%%%%%%%%%%%%%%%%%%%%%%%%%%%%%%%%%%%%%%
To explain the features seen in Fig.~\ref{12} in the red detuned case, we can write the Hamiltonian up to liner terms around $K$-points, $H(k_x,k_y)\approx H(\bm{ K})+(k_x-K_x)\,\partial_{k_x}H(\bm{ K}) + (k_y-K_y)\,\partial_{k_y}H(\bm{ K})$, where $K$ can be any of the two non equivalents $K$-points. After performing the Peierls substitution we arrive at  the following time dependent Hamiltonian:
\begin{eqnarray}
\label{c1}
H(k_x,k_y;t)&=&H(\bm{ K})+(k_x-K_x)\,\partial_{k_x}H(\bm{ K}) \\
\nonumber
             &&+ (k_y-K_y)\,\partial_{k_y}H(\bm{ K}) \\
\nonumber
             &&+2s\,e^{i\Omega t}\sigma_{\scriptscriptstyle{-}}(\bm{ K}) + 2s\,e^{-i\Omega t}\sigma_{\scriptscriptstyle{+}}(\bm{ K}) \,,
\end{eqnarray}
where $\sigma_{\scriptscriptstyle{\pm}}(\bm{ K})=1/2\,[\partial_{k_x}H(\bm{ K}) \pm i\partial_{k_y}H(\bm{ K})]$, which clearly satisfy $\sigma_{\scriptscriptstyle{-}}=\sigma_{\scriptscriptstyle{+}}^\dagger$. Around $K$-points the $\sigma$ matrices are the following (using the base $\{|c\rangle,|v_\tau\rangle,|x_\tau\rangle \}$ from Eqs.~\ref{b1} and~\ref{b2})
\begin{equation}
\label{sigmas}
\sigma_-(\bm{K})\! =\!\left(
\begin{array}{ccc}
0 & 0 & i\,f_1 \\
i\,f_2 & 0 & 0 \\
0 & f_3 & 0
\end{array}
\right),\, 
\sigma_-(\bm{K}')\! =\!\left(
\begin{array}{ccc}
0 & i\,g_1 & 0 \\
0 & 0 & g_2 \\
i\,g_3 & 0 & 0
\end{array}
\right),
\end{equation}
where $f_i$ and $g_i$ ($i=1,2,3$) are real parameters depending on the hoppings given in Appendix A. The differences in these matrices is one of the manifestations of the optical selection rules shown in the Sec.~\ref{OpticalSelectionRulesandMatrixElements}. In the red detuned case (and for a suitable value of $\hbar\Omega$), the Floquet replicas $|c,0\rangle$ and $|v,1\rangle$ can be brought into almost resonance (see Fig.~\ref{19}), and as a result of the interaction with the laser field an important band repulsion appears. Here we give a simple explanation of this effect by using degenerate perturbation theory. To this end it is sufficient to look at the $K$-points only and the Hamiltonian in Eq.~\eqref{c1} is reduced to:
\begin{align}
H(\bm{ K})+2s\,e^{i\Omega t}\sigma_{\scriptscriptstyle{-}}(\bm{ K})+2s\,e^{-i\Omega t}\sigma_{\scriptscriptstyle{+}}(\bm{ K}) \,,
\end{align}
keeping in mind that in the basis chosen the matriz $H(\bm{K})$ is diagonal. The Floquet matrix~\ref{HF} can be constructed and, in a first approximation at least, we can keep only the replicas $m=0$ and $m=1$. The form of this reduced  matrix clearly depends on the $K$ point under examination, as it obvious from Eqs.~\eqref{sigmas}. At the $K$ point we have:
\begin{equation}
\left(
\begin{array}{ccc|ccc}
\mathbin{{\varepsilon_c}{\scriptstyle{+}}}\hbar\Omega & 0 & 0 &   0 & 0 & 2isf_1\\
0 & \mathbin{{\varepsilon_v}{\scriptstyle{+}}}\hbar\Omega & 0 &     2isf_2 & 0 & 0 \\
0 & 0 & \mathbin{{\varepsilon_x}{\scriptstyle{+}}}\hbar\Omega &     0 & 2sf_3 & 0 \\
\hline
0 & -2isf_2 & 0    & \varepsilon_c & 0 & 0 \\
0 & 0 & 2sf_3    & 0 & \varepsilon_v & 0 \\
-2isf_1 & 0 & 0    & 0 & 0 & \varepsilon_x
\end{array}
 \right).
\end{equation}
Here we see that in a  first order approximation the coupling between $|c,0\rangle$ and $|v,1\rangle$ is through the matrix element $2i\,sf_2$. In the exactly degenerate situation ($\varepsilon_c=\varepsilon_v+\hbar\Omega$) the eigenvalues are:
\begin{equation}\label{eigenv}
\varepsilon \approx \frac{\varepsilon_c+\varepsilon_v+\hbar\Omega}{2} \pm 2s\,|f_2|,
\end{equation}
which gives a symmetric shift around the central value $1/2(\varepsilon_c+\varepsilon_v+\hbar\Omega)$ in the replicas $|c,0\rangle$ and $|v,1\rangle$, as seen in Fig.~\ref{12}(a). The result given in Eq.~\eqref{eigenv} remains valid when $|\varepsilon_c-\varepsilon_v-\hbar\Omega|\ll 4s|f_2|$.
At $K'$ point the situation is quite different. The reduced Floquet matrix is now the following:
\begin{equation}
\left(
\begin{array}{ccc|ccc}
\mathbin{{\varepsilon_c}{\scriptstyle{+}}}\hbar\Omega & 0 & 0 &   0 & 2isg_1 & 0 \\
0 & \mathbin{{\varepsilon_v}{\scriptstyle{+}}}\hbar\Omega & 0 &     0 & 0 & 2sg_2 \\
0 & 0 & \mathbin{{\varepsilon_x}{\scriptstyle{+}}}\hbar\Omega &     2isg_3 & 0 & 0 \\
\hline
0 & 0 & -2isg_3     & \varepsilon_c & 0 & 0 \\
-2isg_1 & 0 & 0    & 0 & \varepsilon_v & 0 \\
0 & 2sg_2 & 0    & 0 & 0 & \varepsilon_x
\end{array}
 \right).
\end{equation}  
and since there is no any direct matrix element which link $|c,0\rangle$ and $|v,1\rangle$, there is no shift in this case, as in Fig.~\ref{12}(a).
Similar considerations can be drawn for the blue detuned regime (Fig. {\ref{12}}(b)).  In this case the shift in the three replicas $|c,0\rangle$, $|v,1\rangle$ and $|x,-1\rangle$ is roughly the same at $K$ (for this particular choice of $\hbar\Omega$),  whereas there is no shift at $K'$. Similar to the red detuned case, around the $K'$ we see dynamical gaps much smaller than those in the red detuned case. This is due to the different coupling strength between the conduction and valence or $x$ band.
%\bibliographystyle{apsrev4-1_title}
%\bibliography{usaj_current}
%,Graphene}
%merlin.mbs apsrev4-1.bst 2010-07-25 4.21a (PWD, AO, DPC) hacked
%Control: key (0)
%Control: author (72) initials jnrlst
%Control: editor formatted (1) identically to author
%Control: production of article title (1) required
%Control: page (0) single
%Control: year (1) truncated
%Control: production of eprint (0) enabled
%

\end{document}